\documentclass[twocolumn]{aastex631}

\newcommand{\psat}{$P_\textrm{sat}\,$}
\newcommand{\tcrit}{$T_\textrm{c}\,$}
\newcommand{\rnorm}{$R_\textrm{N}\,$}
\newcommand{\taueff}{$\tau_\textrm{eff}\,$}

\newcommand*{\ra}[2][]{
    \ang[
        math-degree=\textsuperscript{h},
        text-degree=\textsuperscript{h},
        math-arcminute=\textsuperscript{m},
        text-arcminute=\textsuperscript{m},
        math-arcsecond=\textsuperscript{s},
        text-arcsecond=\textsuperscript{s},
        #1]{#2}
}

\let\tablenum\relax

\usepackage{amsmath}
\usepackage{hhline}
\usepackage[utf8]{inputenc}
\usepackage{multirow}
\usepackage{xspace}

\usepackage[binary-units]{siunitx}
  \DeclareSIUnit{\rnormal}{\text{\ensuremath{R_{\textrm{N}}}}}
  \DeclareSIUnit{\sqrthz}{\ensuremath{\sqrt{\text{\hertz}}}}
  \DeclareSIUnit{\parthz}{\pico\ampere/\sqrt{\hertz}}
  \DeclareSIUnit{\ukarcmin}{\micro\kelvin\text{-arcmin}}
  \DeclareSIUnit{\sqdeg}{\text{deg\ensuremath{^{2}}}}
  \DeclareSIUnit{\deg}{\text{deg}}
  \DeclareSIUnit{\arcmin}{\text{arcmin}}
  \DeclareSIUnit\torr{Torr}
  \DeclareSIUnit\icm{icm}

\def\Snospace~{\S{}\,}

\received{June 18, 2021}
\revised{October 18, 2021}
\accepted{November 5, 2021}

\submitjournal{ApJS}

\shorttitle{The SPT-3G Instrument}
\shortauthors{Sobrin et al.}
\graphicspath{{./}{figures/}}

\begin{document}

\title{The Design and Integrated Performance of SPT-3G}

\correspondingauthor{Joshua~Sobrin}
\email{joshuasobrin@uchicago.edu}


\author[0000-0001-6155-5315]{J.~A.~Sobrin}
\affiliation{Department of Physics, University of Chicago, 5640 South Ellis Avenue, Chicago, IL, 60637, USA}
\affiliation{Kavli Institute for Cosmological Physics, University of Chicago, 5640 South Ellis Avenue, Chicago, IL, 60637, USA}

\author[0000-0002-4435-4623]{A.~J.~Anderson}
\affiliation{Fermi National Accelerator Laboratory, MS209, P.O. Box 500, Batavia, IL, 60510, USA}
\affiliation{Kavli Institute for Cosmological Physics, University of Chicago, 5640 South Ellis Avenue, Chicago, IL, 60637, USA}
\author[0000-0001-5868-0748]{A.~N.~Bender}
\affiliation{High-Energy Physics Division, Argonne National Laboratory, 9700 South Cass Avenue., Argonne, IL, 60439, USA}
\affiliation{Kavli Institute for Cosmological Physics, University of Chicago, 5640 South Ellis Avenue, Chicago, IL, 60637, USA}
\author[0000-0002-5108-6823]{B.~A.~Benson}
\affiliation{Fermi National Accelerator Laboratory, MS209, P.O. Box 500, Batavia, IL, 60510, USA}
\affiliation{Kavli Institute for Cosmological Physics, University of Chicago, 5640 South Ellis Avenue, Chicago, IL, 60637, USA}
\affiliation{Department of Astronomy and Astrophysics, University of Chicago, 5640 South Ellis Avenue, Chicago, IL, 60637, USA}
\author[0000-0002-9962-2058]{D.~Dutcher}
\affiliation{Department of Physics, University of Chicago, 5640 South Ellis Avenue, Chicago, IL, 60637, USA}
\affiliation{Kavli Institute for Cosmological Physics, University of Chicago, 5640 South Ellis Avenue, Chicago, IL, 60637, USA}
\author[0000-0002-7145-1824]{A.~Foster}
\affiliation{Department of Physics, Case Western Reserve University, Cleveland, OH, 44106, USA}
\author{N.~Goeckner-Wald}
\affiliation{Department of Physics, Stanford University, 382 Via Pueblo Mall, Stanford, CA, 94305, USA}
\affiliation{Kavli Institute for Particle Astrophysics and Cosmology, Stanford University, 452 Lomita Mall, Stanford, CA, 94305, USA}
\author{J.~Montgomery}
\affiliation{Department of Physics and McGill Space Institute, McGill University, 3600 Rue University, Montreal, Quebec H3A 2T8, Canada}
\author{A.~Nadolski}
\affiliation{Department of Astronomy, University of Illinois at Urbana-Champaign, 1002 West Green Street, Urbana, IL, 61801, USA}
\author[0000-0003-3953-1776]{A.~Rahlin}
\affiliation{Fermi National Accelerator Laboratory, MS209, P.O. Box 500, Batavia, IL, 60510, USA}
\affiliation{Kavli Institute for Cosmological Physics, University of Chicago, 5640 South Ellis Avenue, Chicago, IL, 60637, USA}

% ...




\begin{abstract}

SPT-3G is the third survey receiver operating on the South Pole Telescope dedicated to high-resolution observations of the cosmic microwave background (CMB).
Sensitive measurements of the temperature and polarization anisotropies of the CMB provide a powerful dataset for constraining cosmology.
Additionally, CMB surveys with arcminute-scale resolution are capable of detecting galaxy clusters, millimeter-wave bright galaxies, and a variety of transient phenomena.
The SPT-3G instrument provides a significant improvement in mapping speed over its predecessors, SPT-SZ and SPTpol.
The broadband optics design of the instrument achieves a \SI{430}{\milli\meter} diameter image plane across observing bands of \SIlist{95;150;220}{\giga\hertz}, with \SI{1.2}{\arcmin} FWHM beam response at \SI{150}{\giga\hertz}.
In the receiver, this image plane is populated with 2690 dual-polarization, tri-chroic pixels ($\sim$16000 detectors) read out using a $68\times$ digital frequency-domain multiplexing readout system.
In 2018, SPT-3G began a multiyear survey of \SI{1500}{\sqdeg} of the southern sky.
We summarize the unique optical, cryogenic, detector, and readout technologies employed in SPT-3G, and we report on the integrated performance of the instrument.
\end{abstract}



\section{INTRODUCTION} \label{sec:introduction}

Measurements of the cosmic microwave background (CMB) have been instrumental to the development of $\Lambda$CDM, the standard model of the composition, structure, and evolution of our universe.
Satellite experiments have measured the CMB temperature anisotropy to the cosmic-variance limit at angular multipoles $\ell \lesssim 1600$~\citep{2013ApJS..208...19H, 2020A&A...641A...6P};
and high-resolution, ground-based experiments have extended these measurements to arcminute scales ($\ell \lesssim 10000$)~\citep{2021ApJ...908..199R, 2020JCAP...12..045C}, providing precise measurements of the base $\Lambda$CDM model parameters.
Extracting further information from the CMB at cosmic-variance dominated scales requires measurements of the polarization anisotropy, which provide additional statistical power to constrain cosmological parameters, as well as unique sensitivity to gravitational waves produced during inflation in the early universe.

The polarization of the CMB can be decomposed into $E$-modes and $B$-modes, which correspond to the curl-free and divergence-free components of the polarization field, respectively.
$E$-modes are created from quadrupole temperature anisotropies during the epoch of last scattering~\citep{1997NewA....2..323H}.
Precision measurements of $E$-mode polarization will continue to improve constraints on physics sensitive to the damping tail of the CMB and serve as a powerful consistency test of the $\Lambda$CDM model~\citep{2014PhRvD..90f3504G}.
$B$-modes are generated by gravitational lensing of primordial CMB $E$-modes, galactic and extragalactic emission, and primordial gravitational waves produced during inflation~\citep{1997PhRvD..55.1830Z, 1997PhRvL..78.2058K}.
Distinguishing lensed $B$-modes and foregrounds from primordial $B$-modes requires an instrument with a combination of low noise-equivalent temperature, arcminute-scale resolution, and broad frequency coverage~\citep{2016ARA&A..54..227K, 2016arXiv161002743A, 2021PhRvD.103b2004B}.

Low-noise maps of the microwave sky with arcminute-scale resolution enable a broad range of cosmology and astrophysics research beyond the CMB temperature and polarization power spectra.
Measurements of the lensing-potential power spectrum~\citep{2019ApJ...884...70W} provide an additional way to further constrain $\Lambda$CDM model parameters and extension models~\citep{2020ApJ...888..119B, 2017PhRvD..95l3529S}.
Massive clusters of galaxies can be detected via the Sunyaev-Zeldovich effect, enabling the creation of mass-limited catalogs of clusters, complete out to the redshifts at which they are first formed~\citep{2020AJ....159..110H,2021ApJS..253....3H}.
These cluster samples provide constraints on cosmological parameters independent of those from the primary CMB~\citep{2013JCAP...07..008H, 2016A&A...594A..24P, 2019ApJ...878...55B}.
High-resolution CMB maps also provide a rich sample of emissive point sources, including active galactic nuclei and high-redshift dusty galaxies~\citep{2016A&A...594A..26P, 2020ApJ...893..104G, 2020ApJ...900...55E}.
In addition, time domain analyses of CMB data monitor the sky for millimeter-wavelength transient sources, and CMB surveys are now producing catalogs of strongly flaring transients~\citep{2016ApJ...830..143W, 2020arXiv201214347N, 2021arXiv210306166G}.
Dedicated analyses of CMB maps are also capable of probing exotic, beyond-the-Standard-Model physics.
For example, searches for both static and time-varying birefringence can constrain extensions of the Standard Model involving new axion-like particles~\citep{2017PhRvD..96j2003B, 2020PhRvD.101h3527N, 2020PhRvD.102h3504B, 2021PhRvD.103d2002B}.

In this paper, we describe an instrument designed to perform these measurements: SPT-3G, the third-generation survey receiver installed on the South Pole Telescope (SPT).
The SPT is a 10-meter telescope located at the Amundsen--Scott South Pole Station in Antarctica and optimized to survey the CMB at millimeter wavelengths~\citep{2008ApOpt..47.4418P, 2011PASP..123..568C}.
The South Pole provides an ideal site for millimeter-wave observations due to its altitude (\SI{2835}{\meter}) and exceptionally low precipitable water vapor~\citep{1998SPIE.3357..486R, 2011RMxAC..41...87R}.
Furthermore, the absence of a 24-hour day-night cycle provides relatively stable, low-noise atmospheric conditions and minimal needs for sun-avoidance measures.
The survey receiver exploits the telescope's large primary aperture, using \num{\sim 16000} polarization-sensitive detectors to provide arcminute-scale resolution maps in three observing bands located in the atmospheric transmission windows at \SIlist{95;150;220}{\giga\hertz}.
The main SPT-3G survey covers a \SI{1500}{\sqdeg} area of southern sky that overlaps the survey of the BICEP3/BICEP Array experiment~\citep{2020SPIE11453E..14M}, which has lower resolution and is optimized to detect degree-scale primordial $B$-modes.
\replaced{The order-of-magnitude increase in detectors in SPT-3G---relative to its predecessor instrument, SPTpol~\citep{2012SPIE.8452E..1EA}---is made possible by advances in detector fabrication, readout multiplexing technology, and cryogenic and optical techniques.}{The order-of-magnitude increase in detectors in SPT-3G---relative to its predecessor instrument, SPTpol~\citep{2012SPIE.8452E..1EA}---is made possible by leveraging dual-polarization tri-chroic pixels, a readout system capable of higher multiplexing factors, and a wider throughput telescope design.}
We detail these advances and the integrated, on-sky performance of the SPT-3G instrument in this work.

The organization of this paper is as follows:
\autoref{sec:optics} discusses the optics design, cold-refractive elements, and anti-reflection strategies;
\autoref{sec:cryogenics} describes the design and performance of the cryogenics system, level of optical loading on the detectors, and sub-kelvin assembly;
\autoref{sec:detectors} covers the design, fabrication, and properties of the detectors;
\autoref{sec:readout} details the readout system, with discussions on noise characterization, operable detector yield, detector cross-talk, and data acquisition;
\autoref{sec:observing} provides an overview of the SPT-3G survey footprints and observing strategy, including a description of calibration observations and relative-calibration procedure;
and \autoref{sec:integrated} addresses the integrated, on-sky performance of the SPT-3G instrument, including measurements of its optical efficiency, spectral response, beam response, sensitivity, and polarization calibration.

\section{OPTICS} \label{sec:optics}

\begin{figure*}[ht!]
\includegraphics[trim=0.0in 2.0in 0in 0.5in,clip=true, width=0.48\textwidth]{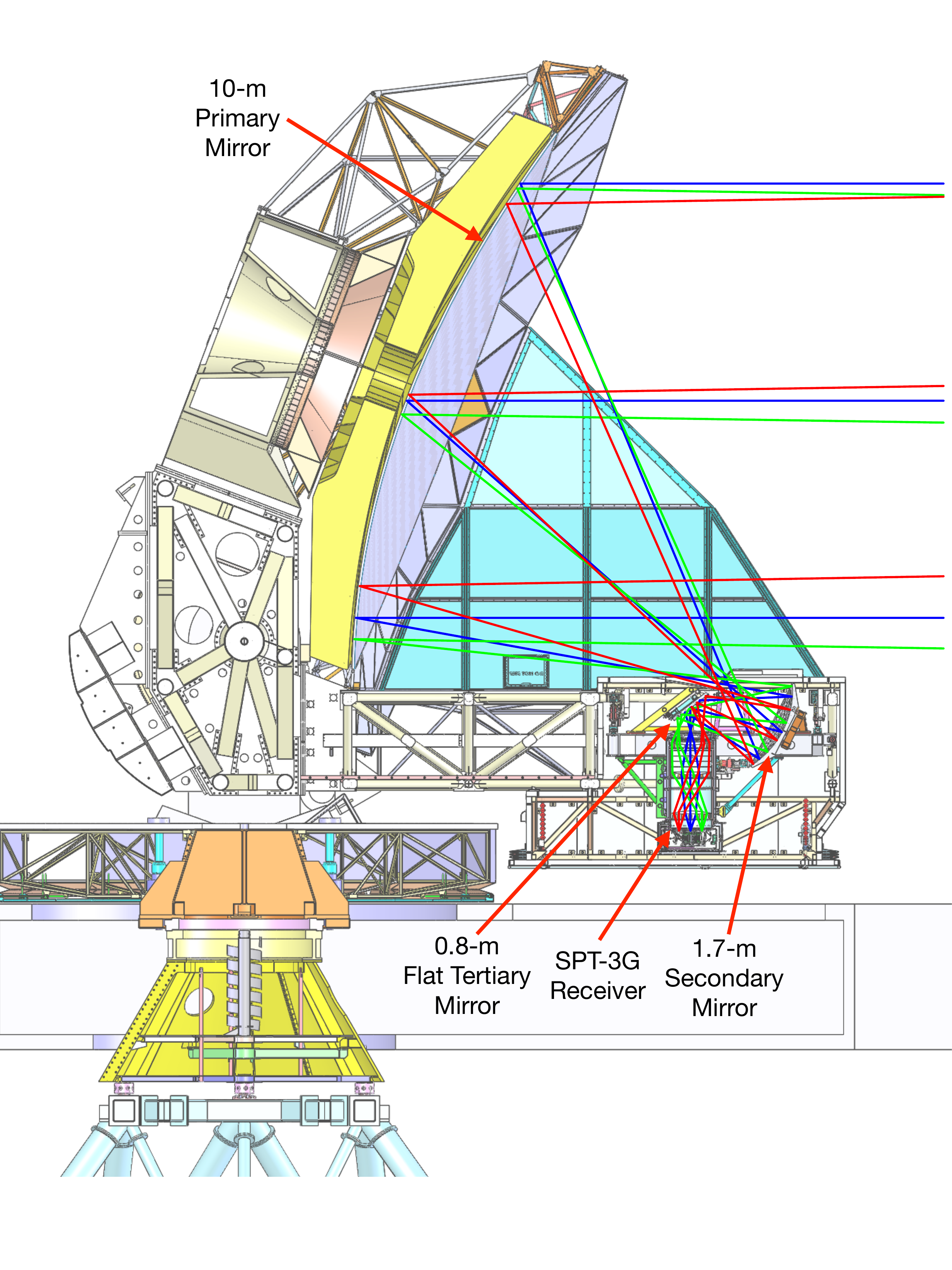}
\includegraphics[trim=-0.3in 0.5in 0in 1.0in,clip=true, width=0.48\textwidth]{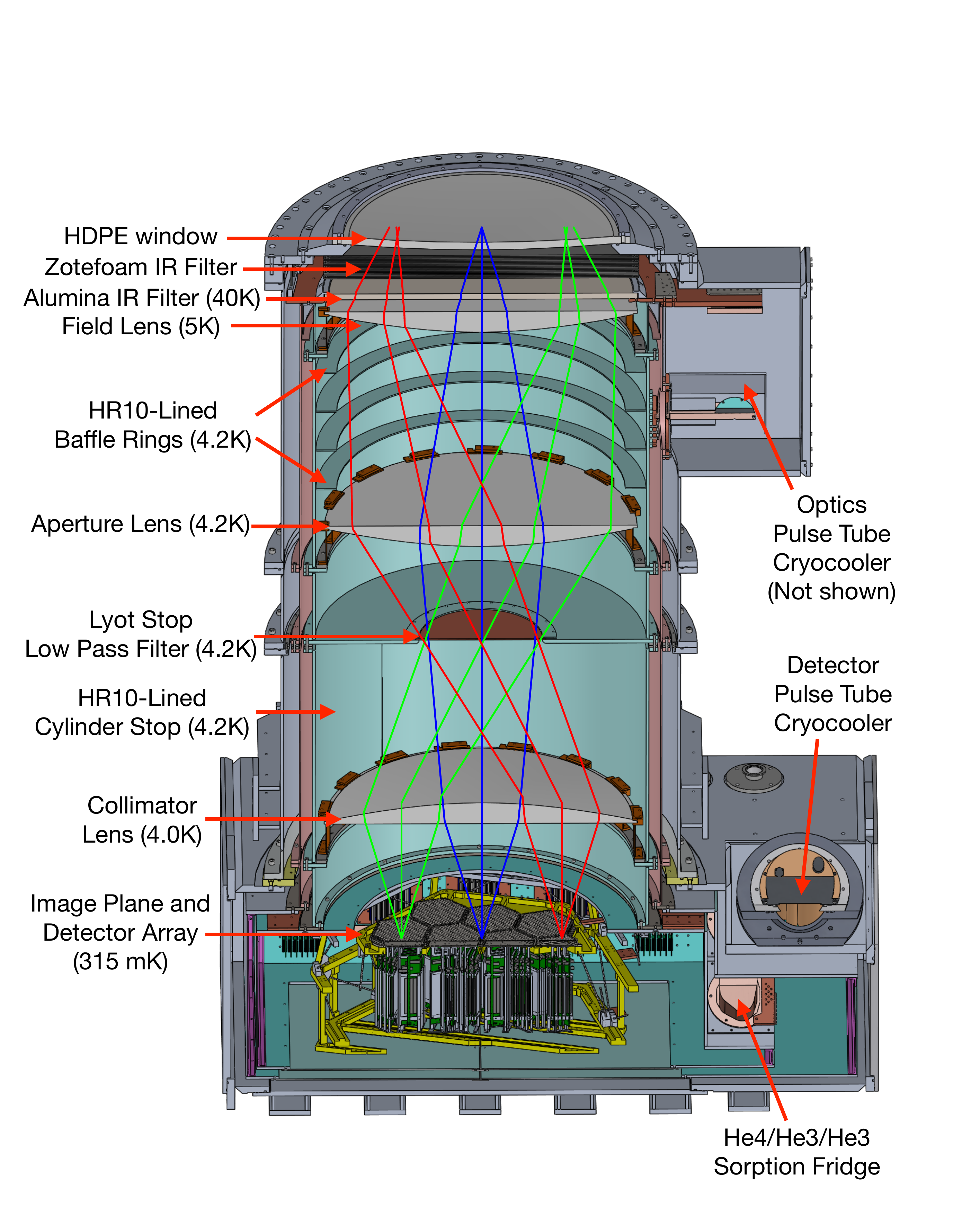}
\caption{
\emph{Left:} Ray trace of the SPT optics design overlaid on a cross-sectional view of the telescope and receiver.
The secondary mirror, tertiary mirror, and receiver are mounted on an optics bench and housed inside a temperature-controlled cabin that moves with the telescope during observations.
Reflected light from the primary mirror passes through an environmental window (\SI{6.35}{\milli\meter} thick HD-30 Zotefoam) on the roof of the cabin.
\emph{Right:} Cross-sectional view of the SPT-3G receiver with ray trace overlaid.
Components of the optics and cryogenics systems are labeled, with typical operating temperatures noted in parentheses.
}
\label{fig:ray_trace}
\end{figure*}

\added{The optics design for SPT-3G aims to maximize the throughput and efficiency of the telescope system while minimizing aberrations, reflections, and scattering.
An annotated illustration of the system is shown in \autoref{fig:ray_trace}.}

For SPT-3G, an off-axis Gregorian design is used to couple the \SI{10}{\meter} diameter primary mirror of the SPT to a new \SI{1.7}{\meter} ellipsoidal secondary~\citep{2018arXiv180908212S} and \SI{0.8}{\meter} diameter flat tertiary mirror.
A set of refractive optics inside the SPT-3G receiver re-images the Gregorian focus onto a flat image plane where the detectors are located.
The primary mirror is under-illuminated to shield against signals from ground emission, sidelobe pickup, and stray light within the cabin and receiver through the use of a \SI{0.28}{\meter} diameter Lyot stop, inside the receiver, which reduces the primary illumination from \SI{10}{\meter} diameter to \SI{8}{\meter} diameter.
The secondary mirror, tertiary mirror, and receiver are mounted on a movable optics bench which can be adjusted relative to the primary mirror to optimize focus quality.
The optics design\deleted{, illustrated in \autoref{fig:ray_trace},} uses pixel sizes of $1.12/ 1.75 / 2.62$~$F\lambda$ at $95 / 150 / 220$~\si{\giga\hertz}.\footnote{$F\lambda$ is defined as the product of the optics design's $f$-number and the wavelength of light.}
In addition to arcminute-scale resolution, this design results in predicted Strehl ratios greater than 0.98 for detectors across all three observing bands over the entire \SI{430}{\milli\meter} diameter image plane and diffraction-limited performance over the \SI{1.88}{\deg} diameter field of view.

\added{Light enters the receiver through a \SI{685}{\milli\meter} diameter vacuum window made of \SI{30}{\milli\meter} thick high-density polyethylene (HDPE).
The AR-coating of the window consists of triangular grooves directly machined into the window surfaces, cut orthogonally on both sides to minimize birefringence and cross-polarization effects~\citep{2016ITTST...6..156S}.
The design of the grooves is based on \citet{Raguin:93}, with groove spacing of \SI{0.652}{\milli\meter} and groove depth of \SI{1.321}{\milli\meter}, which we calculate to result in per-surface reflection of $<$\SI{0.3}{\percent} over the \SI{95}{\giga\hertz} observing band, and $<$\SI{0.1}{\percent} over the \SI{150}{\giga\hertz} and \SI{220}{\giga\hertz} observing bands.}

\added{Beyond the window,} a set of three sintered, poly-crystalline, Coorstek AD-995 aluminum oxide (alumina) lenses comprise the refractive optics chain.
Each lens is a \SI{0.72}{\meter} diameter, plano-convex, 6\textsuperscript{th}-order asphere with center thickness between 54 and 69 \si{\milli\meter}.
Laboratory measurements of Coorstek AD-995 samples indicate an index of refraction of $n = 3.089$ and a loss tangent of  $\tan \delta = 3 \times 10^{-4}$ at \SI{77}{\kelvin}~\citep{2020ApOpt..59.3285N}.

At each pixel of the image plane, an extended hemispherical lens (lenslet) couples radiation to a polarization-sensitive sinuous antenna (\autoref{subsec:pixel_design}).
Lenslets consist of \SI{5}{\milli\meter} diameter alumina hemispheres epoxied to circular depressions in a silicon ``lenslet wafer'' using thin fillets of Stycast 1266~\citep{Nadolski:2020}.
Each array of lenslets is clamped to a silicon wafer containing the antennas and detectors.
The two wafers are positioned relative to each other with an accuracy of $<$\SI{10}{\micro\meter} using alignment marks visible with an infrared microscope.

\begin{figure}[hb!]
\plotone{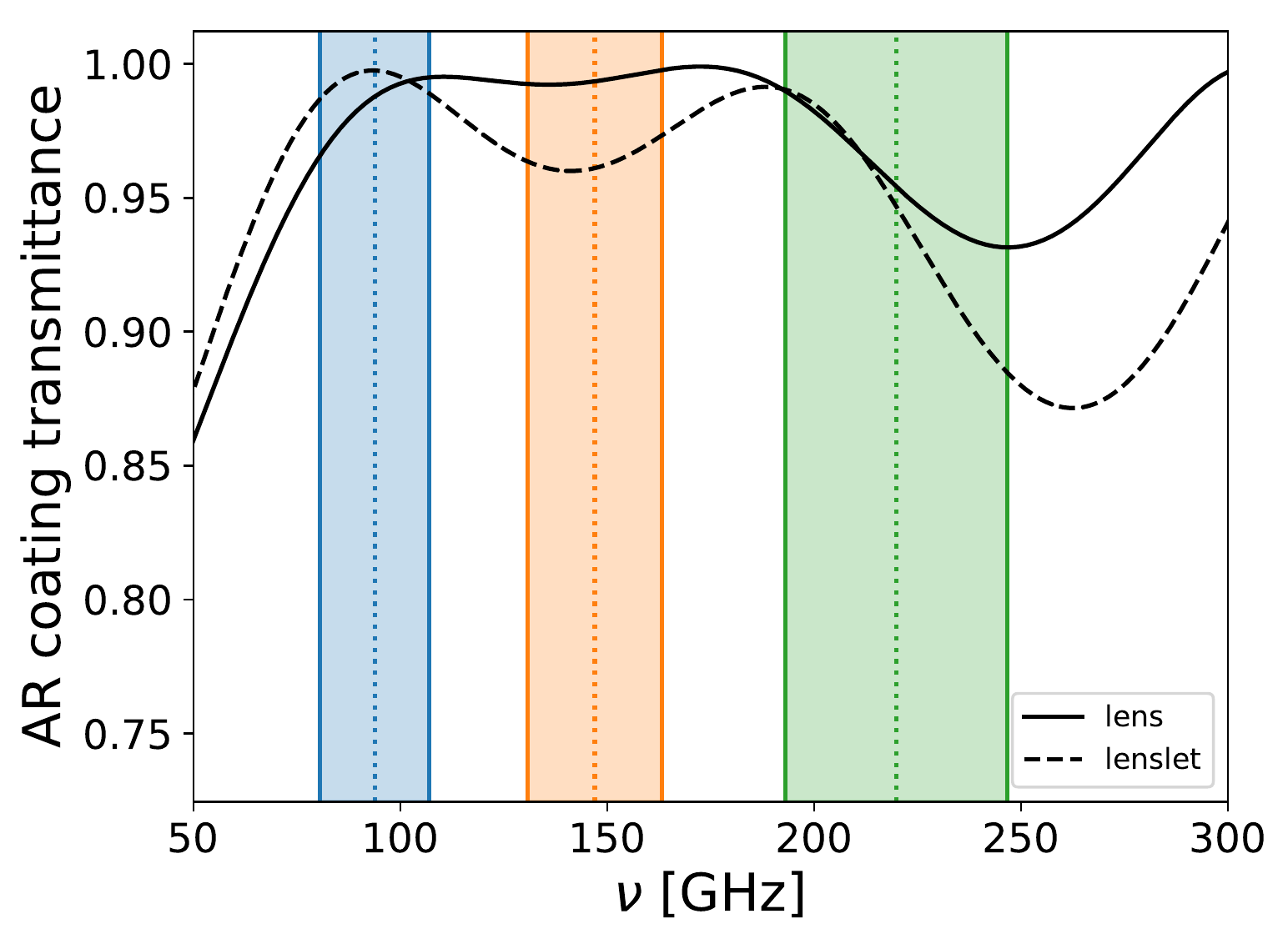}
\caption{
Predicted transmission through a single AR-coated alumina surface, for both lens (solid line) and lenslet (dashed line) prescriptions.
The predictions use laboratory-measured values of dielectric constants and material thicknesses for the constituent materials.
The shaded regions denote the SPT-3G observing bands, centered at \SIlist{95;150;220}{\giga\hertz}~(\autoref{subsec:spectral_response}).
}
\label{fig:ar_transmission}
\end{figure}

Broadband anti-reflection (AR) coatings are used to minimize reflective losses at the alumina lenslets, lenses, and alumina infrared (IR) filter~(\autoref{sec:cryogenics}) across the \numrange[range-phrase = --]{80}{250} \si{\giga\hertz} range.
Separate prescriptions utilizing three distinct layers of polytetrafluoroethylene-based dielectrics were developed for both the large-format monolithic elements and the densely packed lenslet arrays.
A complete description and characterization of these AR coatings can be found in \citet{2020ApOpt..59.3285N}.
\autoref{fig:ar_transmission} shows the expected transmission at lens and lenslet surfaces assuming measured material properties.
At the alumina IR filter and lens surfaces, we expect an average reflection of $ 1.1 / 0.6 / 5.1 $~percent per surface across the $ 95 / 150 / 220 $~\si{\giga\hertz} observing band.
Similarly, at the lenslet surface we expect an average reflection of $ 0.6 / 3.4 / 6.1 $~percent across the $ 95 / 150 / 220 $~\si{\giga\hertz} observing band.
Absorption in the AR coatings is negligible compared to the level of attenuation through the bulk alumina and reflective loss at each surface.

To further control stray light inside the receiver, Eccosorb HR-10 is attached using Stycast adhesive to the innermost aluminum surfaces, both between the collimator lens and Lyot stop and between the Lyot stop and field lens.
HR-10 is flexible, thereby conforming to the aluminum surfaces, and more importantly, has minimal reflection~($\lesssim$\SI{2}{\percent}) even at high angles of incidence~($\lesssim$\SI{80}{\deg}) at millimeter wavelengths~\citep{2010PhDT........61S}.
In addition, a set of four baffle rings covered with HR-10 is positioned between the aperture and field lens to provide surfaces at near-normal incidence to further improve the absorption of stray light.

\section{CRYOGENICS} \label{sec:cryogenics}

\added{The SPT-3G receiver is cryogenically cooled to minimize the instrument's intrinsic in-band thermal emission onto the detector array, as well as to provide a sufficiently cold environment for the detector and readout systems to effectively operate.
In addition, several low-pass optical filters are employed to shield downstream instrument elements and the detector array from sub-millimeter wavelength, out-of-band power.}

The receiver---shown in \autoref{fig:ray_trace}---is functionally separated into an optics and detector cryostat, both cooled by dedicated Cryomech PT-415 cryocoolers (PTCs) while sharing a single vacuum space.
The optics cryostat securely positions and cools the IR blocking filters, lenses, and Lyot stop.
The detector cryostat houses the ten detector modules and cryogenic components of the associated readout electronics.
Before the initial cool-down from room temperature, the receiver is pumped out to $\sim$millitorr pressure over $\sim$24 hours, after which the two PTCs are turned on.
From that point, the receiver needs roughly seven days to reach its base temperature of $\sim$\SI{3}{\kelvin}.

\deleted{Light enters the receiver through a \SI{685}{\milli\meter} diameter vacuum window made of \SI{30}{\milli\meter} thick high-density polyethylene (HDPE).
The AR-coating of the window consists of triangular grooves directly machined into the window surfaces, cut orthogonally on both sides to minimize birefringence and cross-polarization effects~\citep{2016ITTST...6..156S}.
The design of the grooves is based on \citet{Raguin:93}, with groove spacing of \SI{0.652}{\milli\meter} and groove depth of \SI{1.321}{\milli\meter}, which we calculate to result in per-surface reflection of $<$\SI{0.3}{\percent} over the \SI{95}{\giga\hertz} observing band, and $<$\SI{0.1}{\percent} over the \SI{150}{\giga\hertz} and \SI{220}{\giga\hertz} observing bands.}

Past the window, a combination of filters minimizes the transmission of IR radiation, which would raise the base temperatures of the optics elements.
Mounted to the flange holding the vacuum window is a set of ten \SI{3.175}{\milli\meter} thick HD-30 Zotefoam sheets, separated by thermally insulating G-10 spacers.
The Zotefoam sheets are transmissive at millimeter wavelengths and absorptive at IR wavelengths, thereby acting as a multi-layer IR-absorptive filter~\citep{2013RScI...84k4502C, 2018SPIE10708E..2NK}.
We estimate this first filter assembly attenuates any incident IR power through and from the window by a factor of $\sim$10, while attenuating the CMB signal by $<$\SI{1}{\percent}.
Beneath the Zotefoam-filter assembly, and thermally connected to the first stage of the optics PTC, is a \SI{15}{\milli\meter} thick, \SI{720}{\milli\meter} diameter disk of alumina which provides additional IR filtering.
During typical winter observations, this alumina IR filter equilibrates to a temperature of \SI{40}{\kelvin}, with a gradient of $<$\SI{1}{\kelvin} across the filter, thereby contributing a radiative load of $<$\SI{0.1}{\watt} on the first lens.
A final metal-mesh low-pass filter (LPF), with a cutoff of \SI{270}{\giga\hertz}, mounted at the Lyot stop further suppresses power above the observing bands of the detectors~\citep{2006SPIE.6275E..0UA}.

The thermal stages connected to the first- and second-stage cold-heads of the PTCs (henceforth referred to as the 50K stage and 4K stage, respectively), are mechanically supported and thermally isolated by a circular truss assembly of G-10 tubes.
The individual tubes \added{are} \SI{6.35}{\milli\meter} in outer diameter, with \SI{0.76}{\milli\meter} wall thickness, and $\sim$\SI{25}{\milli\meter} length between stages.

\begin{deluxetable}{lcc}
\tablecaption{
Thermal Loading on the SPT-3G 50K and 4K Stages.}
\tablewidth{0pt}
\tablehead{
\colhead{Source} & \colhead{50K Stage} & \colhead{4K Stage} \\
\colhead{} & \colhead{(\si{\watt})} & \colhead{(\si{\milli\watt})} }
\startdata
Vacuum Window & 8.8 & 60 \\
Radiation & 18.1 & 90 \\ 
RF Shielding & 1.5 & 30 \\
Readout Wiring & 2.9 & 130 \\
G-10 Supports & 2.1 & 80 \\
\hline 
Total Predicted Load & 33.4 & 390 \\
\hline
PTC Cold-head Temp. & 29 / 35 \si{\kelvin} & 3.5 / 3.1 \si{\kelvin} \\
Total Inferred Load & 25 & 800 \\
\enddata
\tablecomments{The values for each source are calculated for the fully integrated receiver.
Cold-head temperatures are listed for the optics and detector PTCs, respectively.
Laboratory load-curve measurements are used to infer the actual heat load on each stage from the PTC temperatures.
}
\label{tab:PTC_heat_budget}
\end{deluxetable}

\begin{deluxetable*}{lccc}
\tablecaption{
Predicted Optical Loading and Efficiency for SPT-3G.}
\tablewidth{0pt}
\tablehead{
\colhead{Source} & \colhead{Temperature (\si{\kelvin})} & \colhead{$P_{\textrm{optical}}\phn(\si{\pico\watt})$} & \colhead{Transmission / Efficiency ($\eta$)} \\
\colhead{} & \colhead{} & \colhead{$95\phn/\phn150\phn/\phn220\phn$} & \colhead{$95\phn/\phn150\phn/\phn220\phn$}}
\startdata
Pixel \& Lenslet & 0.3 & $<$0.01 across bands & 0.81 / 0.83 / 0.73 \\
Lyot Stop & 4.2 & 0.16 / 0.10 / 0.01 & 0.56 / 0.82 / 0.96 \\
Metal-mesh LPF & 4.2 & 0.02 / 0.03 / 0.02 & 0.94 / 0.94 / 0.94 \\
3 Lenses & 4--5 & 0.13 / 0.19 / 0.21 & 0.75 / 0.71 / 0.46 \\
Alumina IR Filter & 40 & 0.24 / 0.49 / 0.96 & 0.95 / 0.95 / 0.85 \\
Zotefoam-filter Assembly & $\sim$150--280 & 0.03 / 0.15 / 0.26 & 1.00 / 1.00 / 0.99 \\
Vacuum Window & 280 & 0.73 / 2.59 / 2.50 & 0.98 / 0.96 / 0.95 \\
Cabin Environmental Window & 240 & 0.01 / 0.03 / 0.03 & $>$0.99 across bands \\
Telescope Mirrors & $\sim$250--280 & 0.45 / 1.36 / 1.31 & 0.99 / 0.98 / 0.97 \\
\hline
\textbf{Total Instrument} &  & \textbf{1.78 / 4.94 / 5.30} & \textbf{0.29 / 0.41 / 0.23} \\
$\phn\phn$(excluding stop efficiency) & & ... & 0.52 / 0.50 / 0.24 \\
\hline
Atmosphere & 230 & 1.69 / 2.83 / 1.91 & 0.93 / 0.95 / 0.95 \\
CMB & 2.7 & 0.11 / 0.13 / 0.03 & ... \\
\hline
\textbf{Total} &  & \textbf{3.58 / 7.90 / 7.24} & \textbf{0.27 / 0.38 / 0.22}
\enddata
\tablecomments{
For each element, the optical-power contributions are dependent on the element's temperature, emissivity, reflectivity, scattering, and absorption.
Each element's efficiency is defined as the band-averaged efficiency, averaged over the expected bandpass of the in-line filters on the detector wafers (\autoref{subsec:pixel_design}).
The efficiency values for the pixel \& lenslet are based on laboratory measurements using a tunable blackbody radiative source to illuminate SPT-3G pixels~\citep{2020JLTP..199..320A}.
The Lyot stop efficiencies are the fractions of power that propagate through the stop, estimated from time-reverse Gaussian propagation analyses of the beams at the lenslet; therefore, the stop does not contribute true loss to the system, and we additionally include the total-instrument efficiency without it.   
Transmission through bulk elements is calculated assuming an attenuation of $e^{-\delta k z}$, where $\delta$ refers to the loss tangent of the material, $k$ refers to the wavenumber, and $z$ refers to the thickness of the element.
Loss at the telescope mirrors is dominated by the panel gaps in the primary mirror ($\sim$\SI{1}{\percent} of the total primary area)~\citep{2011PASP..123..568C}, and the finite conductivity of the aluminum used for all three mirrors.
Atmospheric loading is estimated assuming \SI{50}{\percent} median precipitable water vapor in the six month period between June through November at the South Pole, using the AM code~\citep{2019zndo...3406483P}.
The table provides the predicted efficiency of any detector to a single polarization.
}
\label{tab:bolometer_loading}
\end{deluxetable*}

The base temperatures of elements within the receiver are limited by the thermal loading on the PTCs.
To mitigate the development of significant thermal gradients across the receiver, a combination of multi-layer insulation (MLI), high thermal-conductivity material choices, and IR-filtering techniques are employed, as described in \citet{2018SPIE10708E..1HS}.
Elevated base temperatures within the receiver would contribute to increased extraneous optical power on the detectors, a lower observing efficiency, and deteriorated SQUID amplifier performance (\autoref{sec:readout}).
\autoref{tab:PTC_heat_budget} shows the predicted thermal power on both stages from a combination of heat sources.
Radiative sources include out-of-band (primarily IR) thermal power coming through and from the vacuum window and alumina IR filter, along with radiative heat transfer between different temperature-stage cryostat shells.
Conductive sources include the readout wiring, aluminzed-mylar radio-frequency shielding (RF shielding), and G-10 truss assemblies mechanically supporting the radiation shields and alumina IR filter.
The measured temperatures of the PTC cold-heads provide a rough probe of the actual heat load on each thermal stage, from which we infer lower-than-expected and higher-than-expected loading on the 50K and 4K stage, respectively.
Given the lower optics PTC 50K cold-head temperature, it is possible that the vacuum window equilibrates to a colder-than-expected temperature during normal operations, thereby emitting less thermal radiation to the alumina IR filter through the Zotefoam-filter assembly.
It is also possible that the MLI surrounding the entire 50K stage is slightly more effective than predicted.
Considering the higher optics PTC 4K cold-head temperature, we suspect higher-than-expected radiative power on the first alumina lens, either through the alumina IR filter or through unanticipated radiative coupling to room-temperature elements at the top end of the optics cryostat.
Nonetheless, the level of additional loading on the 4K stage is not high enough to noticeably affect the helium condensation (and therefore, performance) of the sub-kelvin sorption refrigerator (\autoref{subsec:sub_kelvin}), or dramatically affect the optical loading on the detectors.
The warmest SQUID amplifiers in the receiver operate at \SI{3.9}{\kelvin}, a temperature that does not result in significantly elevated overall system noise.

Each element of the optics system transmits in-band light with imperfect efficiency and contributes its own extraneous optical power onto the detectors through thermal emission.
The instrument's sensitivity to the CMB depends on both its cumulative optical efficiency and its optical loading onto the detectors, details of which are summarized in \autoref{tab:bolometer_loading}. 
Each element's efficiency is determined by both the level of reflection at the surfaces and the level of absorption through the bulk material (which is a function of material loss tangent and thickness).
Extraneous optical power from the instrument is minimized by cryogenically cooling most elements.
In addition, the cold Lyot stop and absorptive baffling define the outer edges of the beams and terminate grazing reflections to minimize the level of diffusive ray scattering onto hot, emissive surfaces.
Through a comparison of saturation powers of optical and dark detectors~\citep{2018SPIE10708E..1ZD}, we find that the optical loading on the detectors is reasonably consistent with the predicted values in \autoref{tab:bolometer_loading}.
The instrument's optical efficiency is further explored in \autoref{subsec:optical_efficiency}.


\subsection{Sub-kelvin Assembly} \label{subsec:sub_kelvin}

The detector array must be held at an equilibrium temperature of $\sim$\SI{300}{\milli\kelvin} to operate, which is achieved with a custom closed-cycle $^3$He-$^3$He-$^4$He sorption refrigerator fabricated by Chase Research Cryogenics~\citep{BHATIA2000685}.
The refrigerator's ultra-cold cooler (UC) is capable of reaching a base temperature of \SI{265}{\milli\kelvin} under a \SI{4}{\micro\watt} load, and is bolstered by an intermediate cooler (IC) and buffer cooler (1K), which reach \SIlist{330;980}{\milli\kelvin} under \SIlist{20;100}{\micro\watt} loads, respectively.

The sub-kelvin thermal stages thermally isolate and position the \SI{22}{\kilo\gram} detector array at the image plane of the optics system, while maintaining sufficiently cold and stable temperatures by contributing minimal thermal loads on the sorption refrigerator.
Graphlite, a proprietary carbon-fiber reinforced polymer\footnote{\url{https://goodwinds.com/product-category/carbon-fiberglass/carbon/solid-round/}}, is used in a truss configuration to robustly position and support the individual thermal stages.
Graphlite has been previously measured to have a thermal conductivity of \SI{1.8}{\milli\watt\per\meter\per\kelvin} at sub-kelvin temperatures~\citep{2008Cryo...48..448R}, a value we found to be consistent with our own laboratory measurements.
Although the inherent thermal conductivity of Graphlite below \SI{4}{\kelvin} is an order-of-magnitude higher than other commercially available plastics such, as Vespel, its superior strength and stiffness provides a better balance of strength and conductivity in this temperature range.
The Graphlite rods are sanded on their ends, concentrically positioned in tapped-holes of aluminum blocks, and bonded using 3M Scotch-Weld Epoxy Adhesive 2216 Gray.
Prototype joints were tension-tested at liquid-nitrogen temperatures, and consistently survived forces in excess of \SI{7000}{\newton}.

\begin{deluxetable}{lccc}
\tablecaption{
Thermal Loading on the SPT-3G Sub-kelvin Stages.}
\tablewidth{0pt}
\tablehead{
\colhead{} & \colhead{1K Stage} & \colhead{IC Stage} & \colhead{UC Stage}\\
\colhead{Source} & \colhead{(\si{\micro\watt})} & \colhead{(\si{\micro\watt})} & \colhead{(\si{\micro\watt})} }
\startdata
Graphlite Struts      &  39  &  2.1   & 0.06\\ 
Wiring   &  11   &  3.3  & 0.48 \\
RF Shielding         &  21   &  ...  &  0.42\\
Radiation  &  0   &  0   & 0.16\\
\hline 
Total Predicted Load & 71 & 5.4 & 1.12 \\
\hline
Cold-head Temp. & $\sim$\SI{1}{\kelvin} & \SI{312}{\milli\kelvin} & \SI{268}{\milli\kelvin} \\
Total Inferred Load & $\sim$100 & 10 & 4.0 \\
\enddata
\tablecomments{The RF shielding bridges directly between the 1K to UC stages, and does not contribute to the IC stage loading. Radiative loads are expected to be negligible, except at the UC stage, where we expect the detector array to absorb a non-negligible amount of thermal radiation from its \SI{4}{\kelvin} surroundings. Load curves provided by Chase Cryogenics are used to infer the actual heat load on each stage from the measured cold-head temperatures. We believe the discrepancy between prediction and measurement at the IC and UC stages is caused by thermal gradients across the 1K and IC stages exacerbating the conductive loads on the IC and UC stages, respectively.}
\label{tab:mk_heat_budget}
\end{deluxetable}

\begin{figure*}[ht!]
\begin{center}
\includegraphics[trim=0.0in 0.0in 0in 0.0in,clip=true, width=0.64\textwidth]{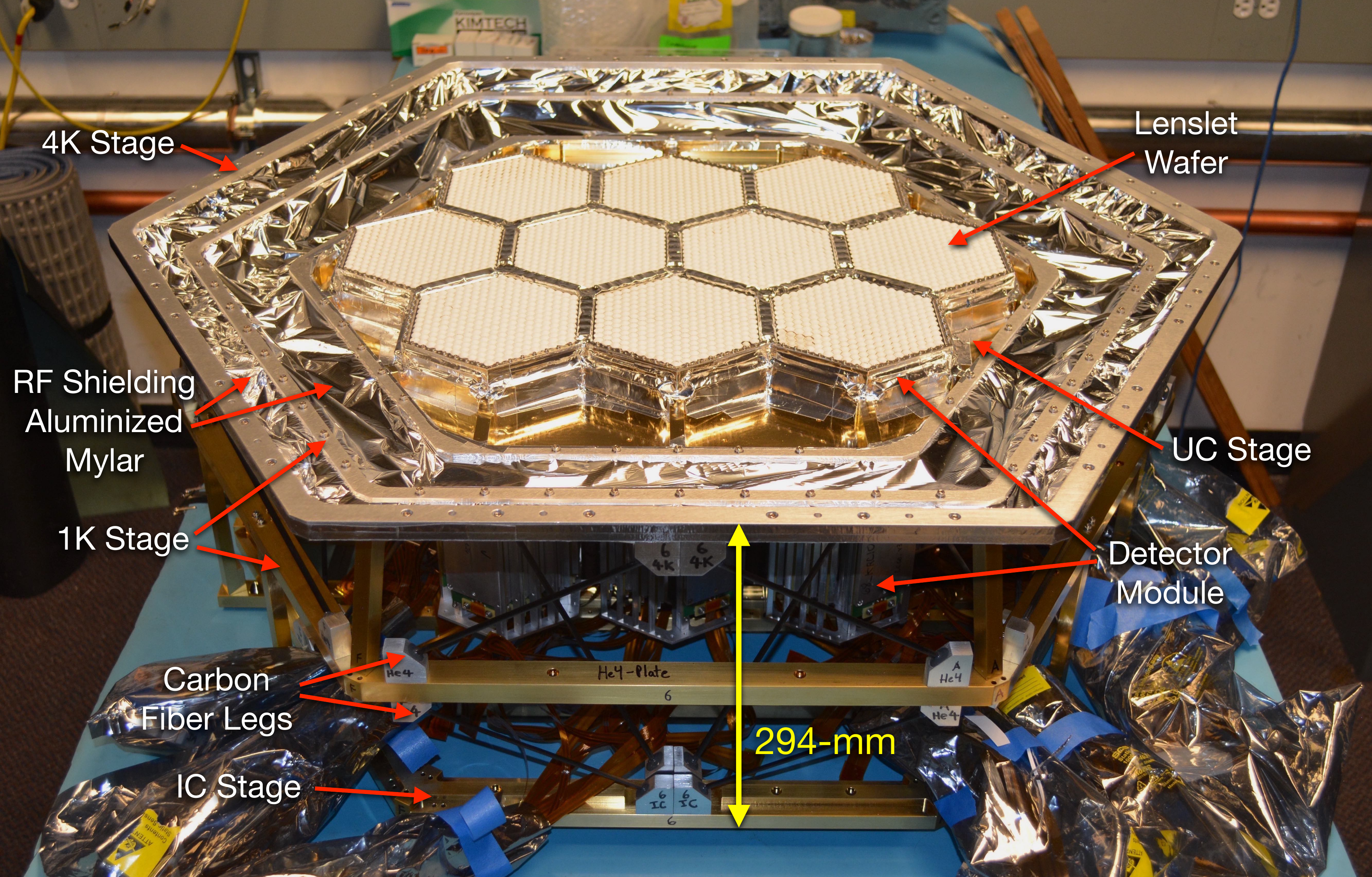}
\includegraphics[trim=-0.0in 0.13in 0in 0.0in,clip=true, width=0.31\textwidth]{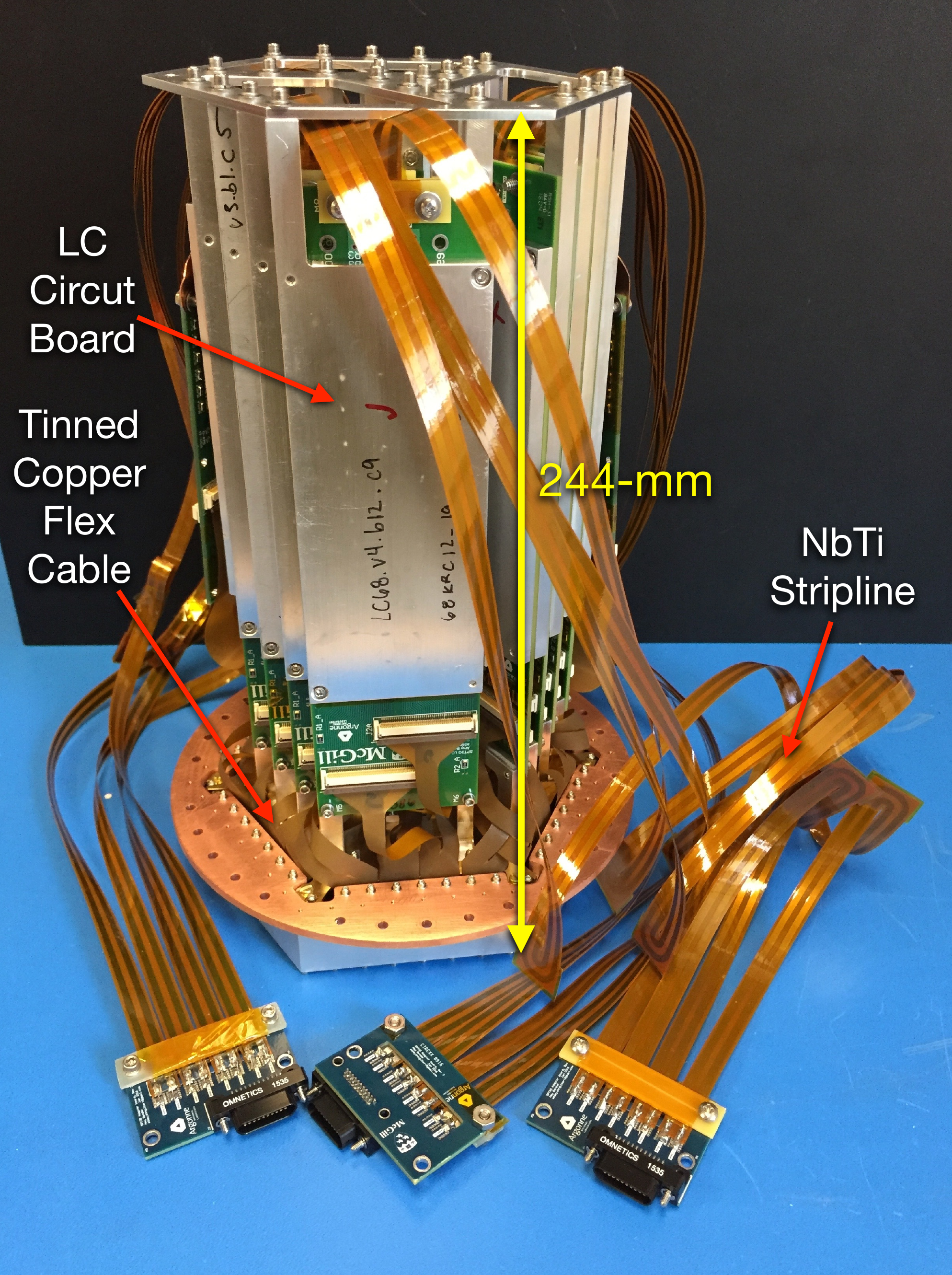}
\caption{
\emph{Left:} Picture of the SPT-3G detector array (consisting of ten detector modules) and supporting sub-kelvin architecture. The thermal stages are mechanically supported by a Graphlite truss structure that stands off the UC, IC, and 1K stages from a mounting ring at 4K.  Each stage is machined from aluminum alloy 6061 and gold-plated to promote thermal conductivity across the components and between interfaces.  The thermal stages are coupled to the sorption refrigerator using pressure-bonded annealed OFHC braided copper straps~\citep{2018SPIE10708E..1HS}.
\emph{Right:}  Picture of a SPT-3G detector module. Each module contains a lenslet array, detector wafer, and twelve LC readout boards. In the picture, the lenslet array is downward facing and hidden by a protective cover. The lenslet array and detector wafer are clamped together in an invar frame, and mounted to a backplate containing Eccosorb AN-72 absorber. The absorber-covered backplate controls the response of the pixels, and also provides a mounting plate for the LC readout boards. Wire-bonded to the detector wafer are flexible cables with tinned-copper traces, which connect to the LC readout boards.  Low thermal-conductance NbTi striplines connect the LC boards to SQUID amplifiers at 4K.}
\label{fig:mK_stage}
\end{center}
\end{figure*}

The detector modules are mounted to the UC stage, which reaches a base temperature of \SI{285}{\milli\kelvin}.
Attached to the back of each detector module is a set of circuit boards that hold the lithographed LC chips (\autoref{sec:readout}).
These boards ($\sim$\SI{300}{\milli\kelvin}) are electrically connected to the SQUID amplifiers ($\sim$\SI{4}{\kelvin}) via low thermal-conductance NbTi striplines, which are heat-sunk to the IC and 1K thermal stages.
This strategy reduces conductive loading at the UC stage by shunting most of the wiring heat to the intermediate thermal stages, which have more cooling power.
An additional source of conductive heat on the 1K and UC stages comes through a continuous sheet of aluminized mylar spanning the gap between the 4K and UC stages.
This sheet is a co-dominant source of loading on the UC stage, but provides an important part of the electromagnetic interference (EMI) shielding around sensitive readout electronics in the detector cryostat.
In combination with a similar sheet between 300K and 4K in the optics cryostat, the aluminized mylar acts as a continuous Faraday cage, shielding the detectors and cryogenic readout electronics from EMI entering primarily through the vacuum window.
The estimated heat loads on each thermal stage are outlined in \autoref{tab:mk_heat_budget}.

This current version of the sub-kelvin assembly, a photo of which is shown in \autoref{fig:mK_stage}, was installed in 2018~December.
An earlier version of the assembly was insufficiently stiff and sensitive to vibrations induced by the telescope's motion, resulting in microphonic heating of the UC stage and excess low-frequency noise in the detectors.
To mitigate these effects, the currently fielded assembly was designed to increase the frequency of the lowest resonant modes while maintaining an acceptable level of conductive heat load between the thermal stages.
Whereas the earlier version utilized oxygen-free high thermal conductivity (OFHC) copper at each thermal stage to minimize thermal gradients, the current design uses thicker gold-plated aluminum at each thermal stage to improve the assembly's stiffness.
Thicker Graphlite struts are now used between the thermal stages, and several design features were added to better account for differential thermal contraction throughout the assembly.
Compared to earlier versions, the current design has a higher total heat capacity and level of heat transfer between thermal stages, thereby decreasing our overall observing efficiency by a few percent.
However, the current version decreases the mass of the total sub-kelvin assembly by \SI{20}{\percent}, and improves the overall stiffness of the assembly by a factor of 2.
We measured the vibrational modes of the assembly and found the lowest mode to be above \SI{50}{\hertz}, consistent with predictions from SolidWorks simulations \added{and well above the expected resonance frequencies of the telescope and optics bench}.
Since the current assembly was installed in 2018, heating of the detector array and excess detector noise due to telescope vibrations are negligible.

During observations, the UC stage is held at \SI{305}{\milli\kelvin} ($\sim$\SI{20}{\milli\kelvin} above the lowest achievable temperature) using a PID controller that adjusts a sorption-pump heater in the refrigerator to control cooling power.
Doing so reduces temperature fluctuations of the detector array to $\lesssim$\SI{0.1}{\milli\kelvin}, slightly improving the hold-time of the refrigerator while negligibly impacting detector performance.
The refrigerator is capable of maintaining the detector modules at this operating temperature for a total of 17 hours before needing to be recycled, which sets a limit on the maximum observing efficiency to $\sim$\SI{80}\%.
Though the refrigerator is designed to support the predicted heat loads for periods of over 72 hours, the achieved hold time is reduced primarily by two issues.
First, the assembly's large heat capacity requires a significant portion of the refrigerator's finite cooling capacity per cycle to be used in cooling the detector array to base temperature from $\sim$\SI{3}{\kelvin}.
Second, given the elevated operating temperatures of the refrigerator IC and UC coolers, we believe that larger-than-expected thermal gradients exist on the 1K and IC stages.
Such gradients would be responsible for larger temperature differentials through conductive components between the thermal stages (e.g. NbTi striplines), leading to higher conductive loads.
This latter issue could be reduced by adding better conductive heat paths across the thermal stages to mitigate gradients, thereby reducing the heat load (and required cooling power) by a factor of $\sim$2 at each stage.
Doing so could potentially increase the time between fridge cycles by $\sim$75\%, but this would only increase the overall, already high, observing efficiency by a few percent.

\section{DETECTORS} \label{sec:detectors}

\begin{figure}[ht!]
\plotone{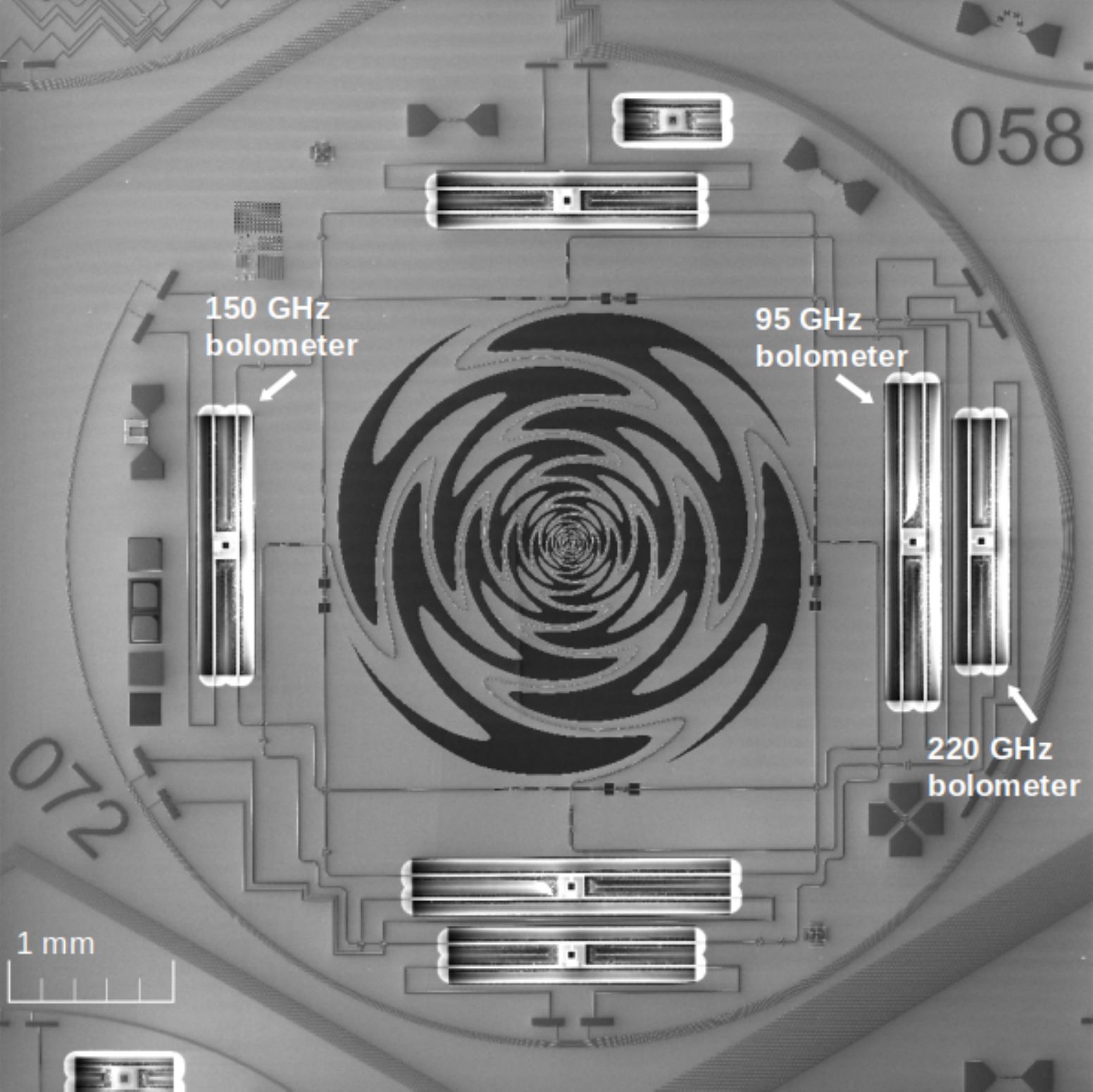}
\caption{
Scanning electron microscope micrograph of an SPT-3G pixel, showing the sinuous antenna at the center surrounded by six TES bolometers as well as various test structures.
The bolometers corresponding to one polarization state have been labeled with their respective observing bands.
}
\label{fig:3g_pixel}
\end{figure}

The SPT-3G detector array includes ten detector wafers, with each wafer containing 269 tri-chroic pixels.
Within each pixel, a broadband dual-polarized antenna couples to transition-edge sensor (TES) bolometers via Nb microstrip transmission lines (\autoref{fig:3g_pixel}).
The signal from the antenna passes through in-line band-defining filters before being transmitted to the respective detectors.
This general pixel architecture was developed at UC Berkeley for POLARBEAR-2/Simons Array~\citep{2016JLTP..184..805S}, and is also planned for use in the Simons Observatory~\citep{2018SPIE10708E..04G} and LiteBIRD~\citep{2018JLTP..193.1048S} experiments.

\begin{figure}[ht!]
\plotone{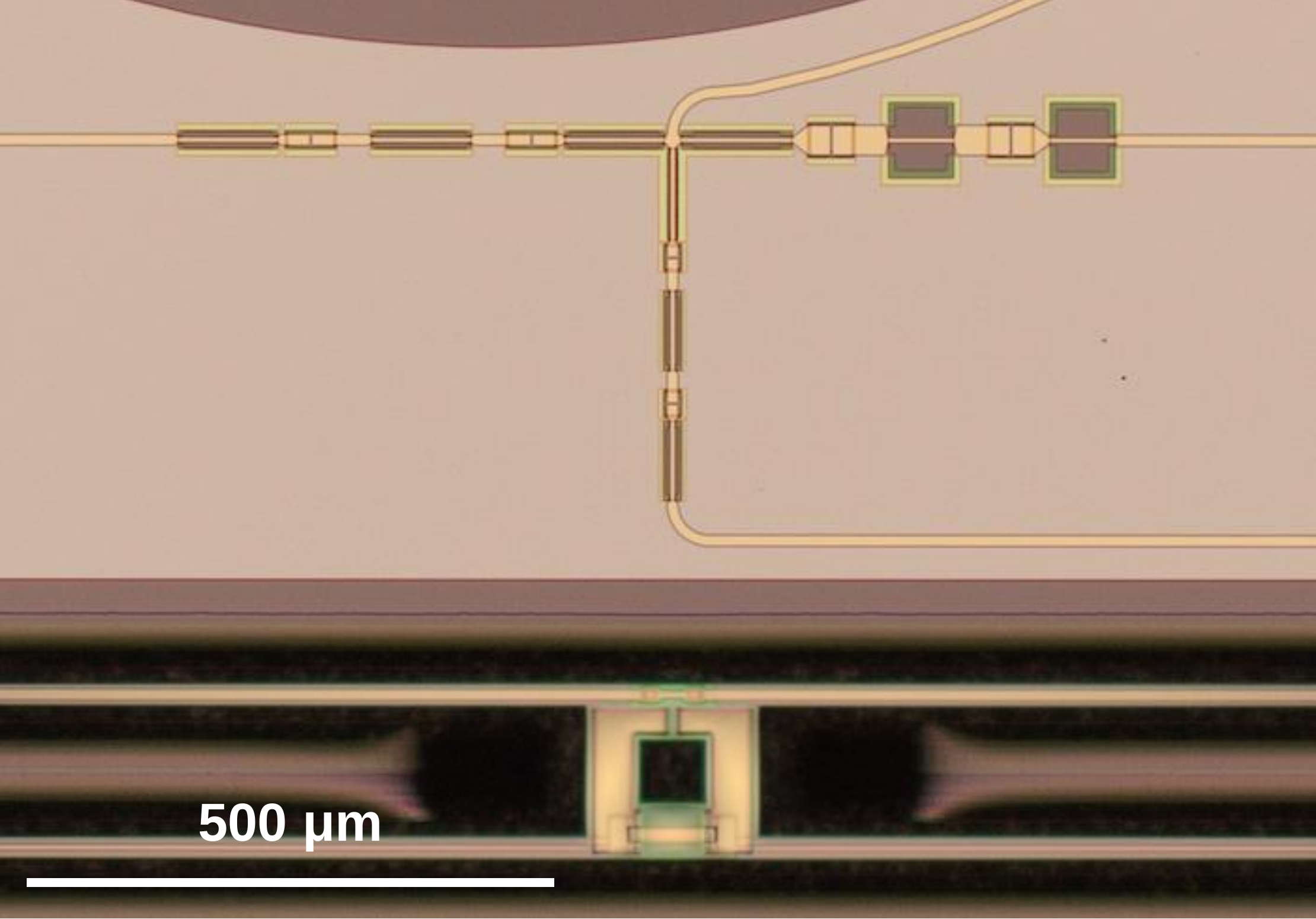}
\caption{
Zoom-in on the SPT-3G pixel, showing a triplexer filter circuit (top) and the bolometer island (bottom).
On the bolometer island, the termination resistor is located at the top while the TES is at the bottom; the largest feature on the island is a layer of palladium that serves as an additional heat capacity.
}
\label{fig:triplexer}
\end{figure}

\begin{deluxetable*}{lccccccc}
\tablecaption{
SPT-3G Detector Properties.
}
\tablewidth{0pt}
\tablehead{
\colhead{} & \colhead{\rnorm} & \colhead{\tcrit} & \colhead{\psat} & \colhead{$G$} & \colhead{\taueff} & \colhead{Readout NET} & \colhead{Total NET}\\
\colhead{} & \colhead{(\si{\ohm})} & \colhead{(\si{\milli\kelvin})} & \colhead{(\si{\pico\watt})} & \colhead{(\si{\pico\watt\per\kelvin})} & \colhead{(\si{\milli\second})} & \colhead{(\si{\micro\kelvin\sqrt{s}})} & \colhead{(\si{\micro\kelvin\sqrt{s}})}\\
\colhead{Wafer} & \colhead{} & \colhead{} & \colhead{95 / 150 / 220} & \colhead{95 / 150 / 220} & \colhead{95 / 150 / 220} & \colhead{95 / 150 / 220} & \colhead{95 / 150 / 220}}
\startdata
w172 & 2.1 & 423 & 11 / 12 / 12 &    99 / 112 / 112 & \phn8 / 10        / 6  &202 / 171 / 881 &516 / 451 / 1598 \\
w174 & 2.2 & 414 & 11 / 14 / 14 &    108 / 151 / 139 &    10 / 10        / 8  &223 / 191 / 1141 &551 / 466 / 1688 \\
\hline
w176 & 2.2 & 493 & 15 / 17 / 18 &    102 / 120 / 120 & \phn5 / \phd5\phn / 4  &383 / 316 / 1385 &851 / 623 / 1962 \\
w177 & 2.1 & 487 & 13 / 15 / 15 &    100 / 116 / 115 & \phn4 / \phd4\phn / 3  &258 / 220 / 1048 &637 / 487 / 1673 \\
\hline
w180 & 2.0 & 460 & ...           & ...                 & \phn7 / \phd7\phn / 5  &201 / 185 / 1177 &551 / 480 / 1811 \\
w181 & 2.0 & 469 & 12 / 14 / 14 &    111 / 124 / 122 & \phn5 / \phd6\phn / 3  &214 / 160 / 926 &651 / 472 / 1558 \\
\hline
w188 & 2.0 & 459 & 11 / 13 / 12 & \phn90 / 109 / 104 & \phn9 / \phd8\phn        / 7  &218 / 187 / 1080 &558 / 469 / 1650 \\
\hline
w203 & 2.6 & ...  & ...           & ...                 & \phn4 / \phd5\phn        / 2  &196 / 139 / 828 &529 / 426 / 1516 \\
w204 & 2.7 & 432 & 11 / 16 / 18 &    103 / 136 / 157 & \phn4 / \phd4\phn / 2  &242 / 198 / 1174 &605 / 482 / 1738 \\
\hline
w206 & 1.8 & 444 & 10 / 13 / 15 & \phn85 / 113 / 128 & \phn5 / \phd4\phn / 2  &169 / 133 / 949 &619 / 484 / 1761 \\
\enddata
\tablecomments{
Median values of the TES normal resistance, \rnorm, transition temperature, \tcrit, bolometer saturation power, \psat, dynamic thermal conductance, $G$, optically-loaded thermal time constant, \taueff, NET from the readout noise per bolometer for each SPT-3G detector wafer, and total NET per bolometer are shown.
Parameters are split by observing band, where pertinent, and horizontal lines indicate wafer fabrication batches.
Some testing data was not available for all wafers.
NET measurements are taken in situ in the deployed instrument, and are described in more detail in \autoref{subsec:noise} and \autoref{subsec:sensitivity}.
}
\label{tab:detector_param}
\end{deluxetable*}


\subsection{Pixel Design} \label{subsec:pixel_design}

At the center of each pixel is a sinuous antenna~\citep{2010SPIE.7741E..0JO}, a type of log-periodic antenna that has several desirable properties, including dual linear polarization, planar geometry which allows for simple lithographic fabrication, low cross-polarization, and broadband response with nearly frequency-independent input impedance.

Sinuous antennas exhibit a periodic variation of polarization angle with frequency, or ``polarization wobble.''
Ansys HFSS simulations of this antenna design show polarization wobble with an amplitude of \SI{\pm 5}{\deg}~\citep{Edwards2012DualPolarizedSA}.
To reduce bias in the polarized maps, an equal number of mirror-image antennas are included on each wafer, such that the biases induced by the left-handed and right-handed wobbles cancel out on average.
In addition to the mirror-image antennas, half of the antennas on each wafer are rotated by \SI{45}{\deg} to evenly sample Stokes $Q$ and $U$ parameters.

The signal from each antenna couples to a microstrip transmission line that lies atop the metalized antenna arms.
An in-line triplexer circuit (\autoref{fig:triplexer}) separates the broadband signal from the antenna into the three observing bands, centered at approximately \SIlist{95;150;220}{\giga\hertz}.
The triplexer~\citep{2012JLTP..167..852S} consists of quasi-lumped-element filters, in which sections of microstrip are removed or formed into high-impedance coplanar waveguides to serve as capacitors and inductors, respectively.
Each microstrip transmission line carrying the signal for a particular polarization and frequency terminates in an impedance-matched \SI{20}{\ohm} resistor on the bolometer island.

The bolometer island is suspended by four silicon nitride legs (\autoref{fig:triplexer}).
The silicon nitride legs carry the antenna transmission and TES bias lines, while also providing a weak thermal link to the bulk of the detector wafer.
The bolometer island contains the termination resistor that thermalizes the signal from the antenna, the TES which measures this temperature change, and an additional heat capacity that serves to increase the bolometer thermal time constant.
The TESs are held in their superconducting transitions by a tunable bias voltage and operated under negative electrothermal feedback, with a loopgain of $\sim$5--10.

\subsection{Detector Fabrication} \label{subsec:detector_fab}

The SPT-3G detector wafers were fabricated at the Center for Nanoscale Materials at Argonne National Laboratory.
A detailed description of the fabrication process can be found in \citet{2015SuScT..28i4002P, 2018JLTP..193..703P}; a brief overview is given here.

The detector wafers start as \SI{675}{\micro\meter} thick, \SI{150}{\milli\meter} diameter silicon wafers coated with low-stress silicon nitride.
The \SI{300}{\nano\meter} thick Nb ground plane layer is deposited and patterned to form the sinuous antennas and basic features of the triplexers and bolometers.
The wafer is heated to \SI{250}{\degreeCelsius} for the deposition of the \SI{500}{\nano\meter} thick SiO$_{\mathrm{x}}$ dielectric layer to ensure a conformal film.
The termination resistors and the TES bolometers are then deposited as thin bimetallic films, consisting either of Ti/Au~\citep{2018JLTP..193..695C} or Al--Mn~\citep{2020JLTP..199..320A}, with similar performance from both types of devices.
The top layer of Nb is then deposited and patterned using a two-step process: lift-off for the TES and resistor leads, and etching for the microstrip and array-level wiring.
An \SI{850}{\nano\meter} thick layer of palladium is deposited on the bolometer islands and partially overlaps the TES to serve as an additional heat capacity.
Lastly, the wafers are diced to their final dimensions, and a XeF$_2$ etch removes the silicon beneath the bolometer islands.

\subsection{Detector Properties} \label{subsec:detector_prop}

The electrothermal properties of the bolometers define their stability and noise performance.
Sources of non-photon noise in the detectors include Johnson noise in the TES bolometers; thermal fluctuations between the bolometers and the bath (or phonon noise); and noise in the readout system.
These noise sources generally increase with detector saturation power, \psat, and so control over this parameter is critical for the instrument performance.
Our target for \psat is twice the expected optical loading values shown in \autoref{tab:bolometer_loading}; this keeps \psat low while enabling stable operation under a range of observing conditions.
Both the TES transition temperature, \tcrit, and the thermal conductance to the bath, $G$, affect \psat, as discussed in \citet{2018JLTP..193..712D}.
To achieve a different \psat for each observing band, \tcrit is held fixed while $G$ is adjusted by altering the bolometer leg lengths.
The TES normal resistance, \rnorm, is the same for all detectors, as its optimal value is constrained predominantly by the readout; see \autoref{sec:readout} for a discussion of readout noise.

The bolometer thermal time constant, $\tau_{\rm eff}$, is set by $G$, the heat capacity of the bolometer island, $C$, and the loopgain of the electrothermal feedback, $\cal{L}$, according to the equation: 
\begin{equation}
\tau_{\rm eff} = \frac{C}{G} \frac{1}{1 + \cal{L}}.
\end{equation}
For stable TES operation, the thermal response time of the bolometers must be slower than that of the feedback circuit, but a fast thermal time constant is desired to preserve the angular resolution of the instrument along the telescope's scanning direction.
These competing requirements restrict $\tau_{\rm eff}$ to lie in the range of $\sim$1--\SI{10}{\milli\second}.
\added{Accurate measurements of $\tau_{\rm eff}$ can be used to deconvolve each detector's temporal response function from time-ordered data before mapmaking for science analyses.}

In \autoref{tab:detector_param}, we give the measured values of selected parameters for the deployed detector wafers.
Wafers were fabricated in batches of five, with occasional changes to target parameters and layer geometries, based on feedback from laboratory testing, leading to some level of variation between batches.
The best performing wafers across batches were then chosen for final installation into the instrument.
Eight detectors wafers were also characterized on the instrument during the 2018 observing season~\citep{2018SPIE10708E..1ZD}, while wafers w204 and w206 were installed in 2018~December to replace two wafers with lower performance.

\section{READOUT} \label{sec:readout}

SPT-3G uses a $68\times$ digital frequency-domain multiplexing (DfMux) architecture~\citep{2014SPIE.9153E..1AB, 2016SPIE.9914E..1DB}.
In this scheme, each bolometer is biased with an AC voltage at a unique frequency between \SIlist{1.6;5.2}{\mega\hertz}, corresponding to the resonant frequencies of a parallel network of LC filters in which the detectors are embedded ($L=\SI{60}{\micro\henry}$, $C=$~\SIrange{14}{148}{\pico\farad}).
Incident radiation from the sky modulates the resistance of each bolometer, and therefore the amplitude of the current flowing through it.
Each LC network is composed of lithographed interdigitated capacitors and spiral inductors on silicon chips, which are mounted behind the detector wafers at the same temperature stage (\autoref{fig:mK_stage}).

After a group of 68 multiplexed channels (henceforth referred to as a ``multiplexing module'') are summed, the signals are carried on broadside-coupled NbTi striplines to NIST SA13 SQUID series arrays ($L_{\textrm{input}} =$~\SIrange{60}{80}{\nano\henry}), mounted at the 4K stage of the detector cryostat~\citep{2011ITAS...21..298S, 211266, 2018SPIE10708E..03B}.
Since SQUIDs are nonlinear amplifiers with limited dynamic range, we use a digital active nulling (DAN) feedback scheme to linearize their performance~\citep{2012SPIE.8452E..0ED}.
DAN uses an integral feedback loop operated by a field-programmable gate array (FPGA) on room-temperature ICE electronics~\citep{2016JAI.....541005B} to null the signals in a narrow band ($\sim$few~kHz) around each detector bias frequency, removing the vast majority of current at the SQUID input coil due to the bolometer bias.
This has the additional benefit of creating a virtual ground before the SQUID input coil, eliminating the impedance of the input coil in series with the bolometer, and therefore improving detector linearity. 
The sky signal is reconstructed by digitally demodulating this nuller signal in a narrow band around each bias frequency.

The readout electronics are designed to achieve detector stability, low readout noise-equivalent power (NEP), and low crosstalk between detectors.
TES bolometers become unstable when the impedance in series with the bolometer is a significant fraction of the detector operating resistance.
This requires minimizing the inductance of the wiring between the LC network and the SQUIDs, motivating the use of low-inductance, low thermal-conductivity NbTi striplines, electrically connected using an ultrasonic soldering system described in \citet{2018JLTP..193..547A}.
In addition, it also defines a minimum operating resistance for the TES bolometers.
Achieving low readout NEP generally pushes the detector resistance in the opposite direction: for a lower-resistance detector, readout current noise terms refer to a smaller power at the detector.
At the same time, the SQUID input inductance and the bolometer resistance act as a current divider for readout noise terms between the SQUID output and integral feedback loop, which imposes a further requirement that the SQUID input impedance be small relative to the bolometer resistance~\citep{2018SPIE10708E..03B}.
Several distinct mechanisms produce electrical crosstalk, which can be mitigated by careful design choices, as described in \autoref{subsec:crosstalk}.


\subsection{System Noise} \label{subsec:noise}

We measure the readout noise in situ by tuning the detectors as usual, then slewing the telescope to the horizon, where the optical power from the atmosphere saturates the bolometers.
This eliminates the photon and phonon noise, leaving only the readout noise (dominant) and TES Johnson noise (sub-dominant) contributions.
The conditions of the measurement modify the observed noise: the lack of detector responsivity no longer suppresses TES Johnson noise, resulting in an increase to this noise source; and the incident atmospheric power raises bolometer resistances out of the transition to the normal resistance, \rnorm, slightly decreasing the observed readout noise.
Each of these effects modify the measured noise by less than \SI{10}{\percent}, and oppose one another, so that noise measured at the horizon is a good approximation of the total readout noise in transition.
In this configuration, the median measured readout noise is $10.4 / 13.0 / 16.0$ \si{\parthz} for $95 / 150 / 220$ \si{\giga\hertz} detectors.
The correlation of readout noise with observing band occurs because readout noise increases with detector bias frequency, and the \SIlist{95;150;220}{\giga\hertz} detectors are arranged in consecutive blocks of increasing bias frequency~\citep{2021arXiv210316017M}.
In \autoref{tab:detector_param}, we compare the readout noise, converted to an effective noise-equivalent temperature (NET), to the total NET from all noise sources.
The contribution of the readout to the total NET is largest at \SI{220}{\giga\hertz}, which reflects the higher intrinsic readout noise, lower detector responsivity (\autoref{subsec:optical_efficiency}), and lower optical efficiency of this band relative to \SIlist{95;150}{\giga\hertz} bands.
Total NET values are measured using the median noise with the detectors in-transition, which is measured daily across the entire 2019 observing season.

\subsection{Crosstalk} \label{subsec:crosstalk}

Electrical crosstalk between bolometers in SPT-3G is predominantly sourced within a multiplexing module.
There are three expected origins of this crosstalk~\citep{2012RScI...83g3113D, 2021arXiv210316017M}.
First, as described above, the resonant bandwidths of LC filters that are nearest neighbors in frequency have nonzero overlap.
This allows a modulation of the resistance of one bolometer to modulate the amplitude of the AC tone biasing the neighbor, producing a crosstalk signal.
Second, the wiring impedance in series with the LCR$_{\textrm{TES}}$ network creates a divider for the applied voltage biases.
As the bolometer resistances modulate in response to sky signals, the voltage ratio of this divider is similarly modulated.
Last, pairs of planar spiral inductors have a nonzero mutual inductance, which can result in crosstalk. 

Dense observations of RCW38 (\autoref{subsec:calibration}) are used to measure the crosstalk in SPT-3G.
In these data, the telescope is rastered such that every detector in the focal plane sees the source, and individual maps are made for each detector.
A template model for RCW38 is used to extract the baseline amplitude for the crosstalk ``source'' detector and then also at the positions of the other ``recipient'' detectors within that multiplexing module~\citep{2020JLTP..199..182B}.
Crosstalk components from source detectors within a \SI{5}{\arcmin} radius of the recipient RCW38 centroid are excluded to prevent confusion due to the extended source profile.
Taking the inverse variance weighted average across 28 observations, we find that \SI{89}{\percent} of the crosstalk components meet our design target of \SI{<0.5}{\percent}.
The slight excess of detector pairs with crosstalk \SI{>0.5}{\percent} is correlated with channels whose LC resonant frequencies scattered closer to their nearest-frequency neighbor than the design.
No crosstalk is detected when the analysis is extended to include recipient detectors in other multiplexing modules, indicating that crosstalk is primarily within a module.
Due to the low overall instrument crosstalk, we find no need to analytically remove the measured crosstalk from our data, as we have done for past experiments~\citep{2018ApJ...852...97H}.
The effect of any residual cross talk is accounted for in the calibration.


\subsection{Data Acquisition} \label{subsec:daq}

The detector data are digitized and sampled at \SI{152.5}{\hertz}, packetized by the FPGAs in the ICE readout electronics, and streamed over gigabit ethernet.
These data are serialized and written to disk at a rate of about \SI{20}{\mega\byte\per\second} by the data acquisition (DAQ) software (part of the \texttt{spt3g\_software} package~\citealt{spt3gsoftware2019}) running on a computer on the readout network, using the \texttt{cereal} library~\citep{cereal2017}.
The DAQ also mediates the transfer of housekeeping information from the readout electronics and telescope information from the General Control Program (GCP).
GCP is an independent process that handles the telescope control and pointing, and is sampled at a lower rate than detector data~\citep{2012SPIE.8451E..0TS}.

After each observation, detector data are compressed by a factor of $\sim$8 using the lossless FLAC compression algorithm\footnote{\url{https://xiph.org/flac}} and merged with the housekeeping, calibration, and pointing information to form the raw data that is the input to the data analysis and mapmaking pipeline.
The SPT-3G observing cadence results in about \SI{300}{\giga\byte} of compressed data stored to disk daily.
To facilitate data transfer from the South Pole for timely data analysis, we further downsample the time-ordered data by a factor of two and remove the demodulator quadrature that is out-of-phase with the bolometer response.
These downsampled data are transferred every day via the TDRS satellite network, \added{and are primarily used for analyses that do not require high resolution ($\lesssim$\SI{2}{\arcmin}) maps.}
In addition to the full downsampled data set, about half of the full-rate (compressed) data are also transferred north via TDRS each day; the rest of the full-rate data remain on local storage at the South Pole until they can be shipped out during the austral summer season each year.
Some online data processing is performed on the computing system at the South Pole in near real-time, such as analysis of calibration data, preliminary mapmaking for monitoring data quality, and a transient alert pipeline.

\section{OBSERVING STRATEGY} \label{sec:observing}


\subsection{CMB Field Observations} \label{subsec:cmb_fields}

\begin{figure}[ht!]
\plotone{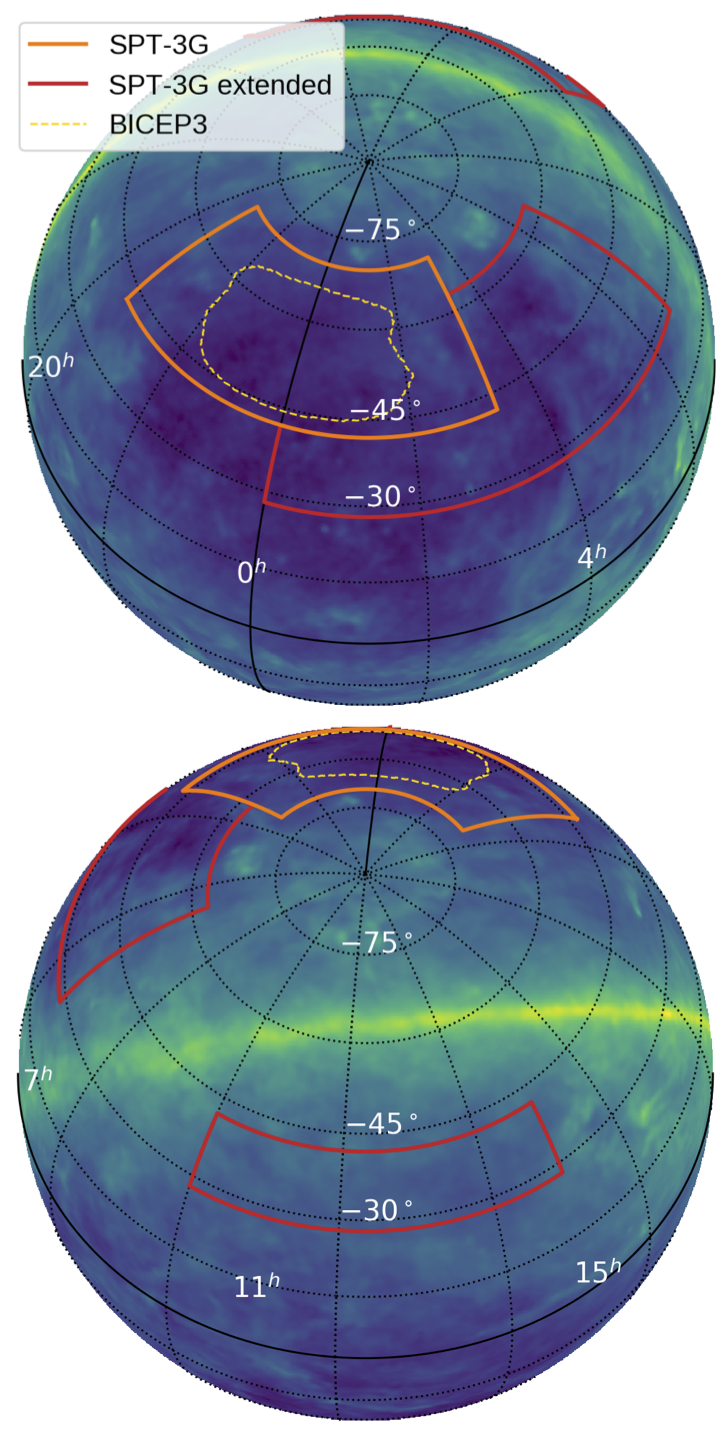}
\caption{
Footprints of the SPT-3G \SI{1500}{\sqdeg} survey \emph{(orange solid)}, the SPT-3G extended survey \emph{(red solid)}, and BICEP3 survey~\citep{2014SPIE.9153E..1NA} \emph{(yellow dotted)}, overlaid on the Planck thermal dust map~\citep{2016A&A...594A..10P}.
The BICEP Array~\citep{2020SPIE11453E..14M} survey area (not shown) is expected to have a similar footprint to BICEP3.
The yellow contour is chosen to correspond to an effective sky fraction at the mean survey weight. 
Since BICEP3 has a much larger field of view than SPT-3G, there is significant sky area outside the yellow contour, but still overlapping with the SPT-3G survey, on which BICEP3 has nontrivial survey weight.
}
\label{fig:survey_footprint}
\end{figure}

The SPT-3G main survey covers a \SI{1500}{\sqdeg} footprint extending from \ra{20;40;0} to \ra{3;20;0} right ascension (RA), and \ang{-42;;} to \ang{-70;;} declination (dec).
Our choice of survey footprint, shown in \autoref{fig:survey_footprint}, is motivated by the need for high-resolution, low-noise CMB maps to remove the $B$-modes induced by gravitational lensing, which contaminate searches for inflationary degree-scale $B$-modes in BICEP/\emph{Keck} data~\citep{2018PhRvL.121v1301B, 2021PhRvD.103b2004B}.
As a result, the SPT-3G footprint was chosen to closely match that of BICEP Array~\citep{2020SPIE11453E..14M}.
From early December to sunset (March 21), the sun produces a detectable signal in the main survey field because of diffraction sidelobes from panel gaps in the primary mirror~\citep{2012SPIE.8452E..1FG}.
Starting in the 2019-2020 austral summer, and continuing in the 2020-2021 austral summer, we began a summer-only extended survey that extends to both higher RA and higher dec, similar to \citet{2020ApJS..247...25B} (see ``SPT-3G extended'' in \autoref{fig:survey_footprint}).
The SPT-3G extended survey will provide a larger sample of galaxy clusters and improved cosmological constraints because of its increased sky fraction.

The main survey footprint is divided into four subfields centered at \SIlist{-44.75;-52.25;-59.75;-67.25}{\degree} dec, respectively, to limit the variation in detector responsivity throughout each observation.
Similar to previous SPT surveys (e.g. \citealt{2018ApJ...852...97H}), we observe the CMB by rastering the telescope across a subfield at constant elevation, taking a \SI{12.5}{\arcmin} step in elevation, then repeating until the full elevation range of the subfield has been observed, taking a total of $\sim$2 hours.
The raster scans composing each subfield observation have a small global elevation offset of $N \times \SI{0.5}{\arcmin}$, with $0 \leq N < 25$, known as ``dither steps,'' improving the uniformity of coverage in coadded maps.

\subsection{Calibration Observations and Relative Calibration Procedure} \label{subsec:calibration}

In addition to observations of each CMB subfield, a suite of calibration observations is regularly performed to convert the time-ordered data into units of CMB blackbody temperature.
This conversion depends on the optical power incident on each detector, which varies due to changes in weather and the fact that SPT-3G observes regions of sky spanning elevations from \ang{28;;} to \ang{70;;}.
While the final temperature calibration of SPT maps is performed by cross-correlating with Planck maps (\autoref{subsec:beams}), this initial time-domain calibration is used to weight and coadd individual observation maps.

The first step in the calibration chain entails a dense raster scan of one of the Galactic HII regions RCW38 or MAT5a (NGC~3576)~\citep{2003astro.ph..1599C}, in which every detector scans over the source (henceforth referred to as a ``dense observation'').
Both HII regions are observed in this manner every eight days.
RCW38 is used for the calibration of the two lower-elevation subfields of the \SI{1500}{\sqdeg} survey, while MAT5a is used for the two higher-elevation fields.
The HII regions have known reference flux from previous Planck-calibrated observations by SPT-SZ~\citep{2019JCAP...07..038M}.
A map for each bolometer in a dense observation is fit to a band-averaged template, and the fitted source flux is compared to the reference flux to obtain the calibration of each detector in CMB temperature units.
Since the dense observations are taken infrequently, two corrections must be applied to obtain accurate relative temperature calibrations for each CMB field observation.

First, a detector's responsivity may be different during a CMB observation than during the most recent dense observation, due to changes in optical power incident on the detector or detector voltage bias.
A chopped thermal source is located behind a small aperture in the secondary mirror and illuminates all detectors in the focal plane.
The response to this source is measured both immediately before the dense observation, as well as before every CMB field observation, and the ratio of these two responses is used to correct for changes in the detector responsivity.

A second correction accounts for changes in atmospheric transmission between the time of the dense observation and a CMB field observation.
Immediately before every CMB field observation, we perform a faster, sparse raster scan of one of the HII regions, in which only a subset of detectors scans over the source (henceforth referred to as a ``sparse observation''), and we also measure the response of the detectors to the chopped thermal source.
The time-ordered data for the sparse observation are calibrated in units of watts at the TES and corrected for the difference in detector responsivity between the sparse and dense observations as described in the previous paragraph.
For each band, we form a coadded map of these time-ordered data from the sparse observation, and another coadded map from the most recent dense observation of the same HII region.
The band-averaged atmospheric transmission $T_\nu$, at the time of the sparse observation $t_\textrm{sparse}$, relative to the dense observation at $t_\textrm{dense}$, in each observing band ($\nu \in \{ \SIlist[list-separator={, }, list-final-separator={, }, list-units=single]{95;150;220}{\GHz} \}$), is then defined as the ratio of integrals of the maps over a $\SI{4}{\arcmin} \times \SI{4}{\arcmin}$ box:
\begin{equation}
T_\nu(t_\textrm{sparse}) \equiv \frac{\int_{4^\prime\times4^\prime} M_\nu(\Omega; t_{\textrm{sparse}})~d\Omega}{\int_{4^\prime\times4^\prime} M_\nu(\Omega; t_{\textrm{dense}})~d\Omega},
\end{equation}
where $M_\nu$ is the coadded map of the sparse or dense observation of the HII region.
Note that coadded maps are constructed from a subset of bolometers in the focal plane, which observe the chopped source and HII region with high signal-to-noise.

Taking these effects together, $C_i$, the conversion from electrical power at the TES to CMB temperature for bolometer $i$ is expressed as
\begin{align}
\begin{split}
C_i(t_{\textrm{CMB}}) \left[ \textrm{W} / \textrm{K} \right] = &\frac{R_i(t_{\textrm{CMB}}) \left[ \textrm{W} \right]}{R_i(t_{\textrm{dense}}) \left[ \textrm{W} \right]} \times T_{\nu(i)}(t_\textrm{sparse}) \\
&\times \frac{\hat{A}_i \int_{4^\prime\times4^\prime} M_{\nu(i)}(\Omega; t_{\textrm{dense}}) \left[ \textrm{W} \right]~d\Omega}{\int_{4^\prime\times4^\prime} M^\textrm{ref}_{\nu(i)}(\Omega) \left[ \textrm{K} \right]~d\Omega},
\end{split}
\end{align}
where $R_i(t) \left[ \textrm{W} \right]$ is the response of bolometer $i$ to the chopped thermal source at time $t$ in units of watts at the TES, with $t_{\textrm{CMB}}$ and $t_{\textrm{dense}}$ being the times of a CMB observation and its preceding dense HII region observation, respectively;
$M^\textrm{ref}_{\nu(i)}(\Omega) \left[ \textrm{K} \right]$ is the reference map from SPT-SZ of the HII source calibrated in units of $\textrm{K}_\textrm{CMB}$;
and $\hat{A}_i$ is a best-fit amplitude parameter obtained by a linear least-squares fit of a dense HII region map for bolometer $i$ to a template $M_{\nu(i)}(\Omega; t_{\textrm{dense}})$ constructed from a map coadded across a set of detectors.
Coadded maps created using this calibration have an absolute calibration that is within \SI{~10}{\percent} of that of Planck, with nonidealities caused by differences in the bandpasses and beams between SPT-3G and SPT-SZ.

\section{INTEGRATED PERFORMANCE} \label{sec:integrated}


\subsection{Detector and Readout Yield} \label{subsec:yield}

Several stages of characterization during integration and commissioning defined the set of operable detectors.
The first of these was a room-temperature continuity check at the wafer, which identified TES channels that were open, shorted to their neighboring detectors, or shorted to ground, primarily due to a combination of wafer fabrication and wirebonding defects.
The wirebonds of channels shorted to ground were removed to prevent the entire multiplexing module from being shorted to ground.
In total, 14166 out of 15720 detectors (\SI{90.1}{\percent}) tested passed the room-temperature continuity check.

After cooling the receiver to its operating temperature, we performed a network analysis by sweeping a voltage tone across the bandwidth containing LC multiplexer resonances and recording the resulting current through the system, thereby mapping out resonances corresponding to known detectors.
We identified valid resonances for 14261 detectors (\SI{90.7}{\percent}), a higher yield than our warm continuity check because many channel-to-channel shorts at room temperature remain \SI{>1}{\kilo\ohm} at cryogenic temperatures.
We further pruned the set of operable detectors with additional network analyses performed under a \SI{300}{\kelvin} optical load, above and below the TES critical temperature; the resonance shapes under these conditions can indicate detectors that are insensitive to incident radiative power (e.g. due to defects in the TES fabrication). This was the most significant cut, removing 1718 potential detectors.
Additionally, ten full multiplexing modules were disabled because of elevated SQUID or readout noise, or because shorts to ground on the detector array generated thermal heating when operated (660 detectors).
After all cuts, we operated a median of 11424 detectors (\SI{72.7}{\percent}) during the 2019 observing season.


\subsection{Optical Efficiency} \label{subsec:optical_efficiency}

The overall sensitivity of the experiment is dependent on the cumulative optical efficiency of the instrument, which we characterize using the same single-detector maps of RCW38 described in \autoref{subsec:crosstalk} and \autoref{subsec:calibration}.
These single-detector maps are compared against the known brightness temperatures of RCW38 in each band, yielding a cumulative optical efficiency measurement, $\eta$, for every detector.

To calculate $\eta$, the integrated flux across the source is measured by each detector in units of \si{\watt \cdot \steradian}, using the voltage and current calibrations from the readout electronics.
Coadded maps of RCW38 in units of K\textsubscript{CMB}, which are absolutely calibrated against Planck~\citep{2019JCAP...07..038M}, are then used to infer the integrated flux that would be measured by a perfectly efficient polarization-sensitive detector and receiver with top-hat spectral response with band centers and widths as defined in \autoref{subsec:spectral_response}.
The ratio of these two quantities provides a measurement of $\eta$.

This measurement is limited by any systematic errors in accurately converting between a change in measured current at the TES and a change in the optical power deposited at the TES.
These power calibration errors are largely due to small fluctuations in the parasitic impedances in the readout circuit, which are a complicated function of bias frequency and not very well constrained.
Specifically, we define $\eta_{\mathrm{fmux}}$ as the quantity obtained from the procedure described above, which we claim is related to $\eta_{\mathrm{true}}$ through
\begin{equation}
\eta_{\mathrm{fmux}} = \frac{1}{2} V_{\mathrm{fmux}} \frac{dI_{\mathrm{fmux}}}{dI_{\mathrm{true}}} \frac{dI_{\mathrm{true}}}{dP_{\mathrm{true}}} \eta_{\mathrm{true}},
\end{equation}
where $V_{\mathrm{fmux}}$ is the measured voltage bias across the TES, the first derivative term describes the readout system's transfer function when measuring current across the TES, and the second derivative term is the true responsivity of the detector under an AC bias~\citep{Irwin2005}.

\begin{deluxetable}{cccc}
\tablecaption{Measured SPT-3G Optical Efficiency, $\eta$.}
\tablewidth{0pt}
\tablehead{
\colhead{Wafer} & \colhead{\SI{95}{\giga\hertz}} & \colhead{\SI{150}{\giga\hertz}} & \colhead{\SI{220}{\giga\hertz}}
}
\startdata
w172 & $ 0.21 \pm 0.04 $ & $ 0.35 \pm 0.10 $ & $ 0.11 \pm 0.02 $ \\
w174 & $ 0.24 \pm 0.04 $ & $ 0.37 \pm 0.07 $ & $ 0.11 \pm 0.04 $ \\
w176 & $ 0.21 \pm 0.03 $ & $ 0.36 \pm 0.06 $ & $ 0.13 \pm 0.03 $ \\
w177 & $ 0.23 \pm 0.03 $ & $ 0.39 \pm 0.05 $ & $ 0.14 \pm 0.02 $ \\
w180 & $ 0.29 \pm 0.05 $ & $ 0.46 \pm 0.07 $ & $ 0.12 \pm 0.03 $ \\
w181 & $ 0.24 \pm 0.05 $ & $ 0.49 \pm 0.12 $ & $ 0.16 \pm 0.04 $ \\
w188 & $ 0.29 \pm 0.04 $ & $ 0.48 \pm 0.12 $ & $ 0.14 \pm 0.03 $ \\
w203 & $ 0.25 \pm 0.04 $ & $ 0.49 \pm 0.10 $ & $ 0.14 \pm 0.04 $ \\
w204 & $ 0.31 \pm 0.08 $ & $ 0.50 \pm 0.10 $ & $ 0.13 \pm 0.03 $ \\
w206 & $ 0.32 \pm 0.09 $ & $ 0.59 \pm 0.25 $ & $ 0.17 \pm 0.05 $ \\
\hline
\textbf{Full Array} & $ 0.25 \pm 0.07 $ & $ 0.44 \pm 0.14 $ & $ 0.13 \pm 0.04 $ \\
\hline
\textbf{Predicted} & $0.27$ & $0.38$ & $0.22$ \\
\enddata
\tablecomments{
13 calibration observations during the 2019 austral winter were used to determine the median $\eta$ of every operating detector.
These values were used to determine the medians and standard deviations across each wafer, as well as across the full array. 
The predicted efficiencies are based on the model presented in \autoref{tab:bolometer_loading}.
}
\label{tab:optical_efficiencies}
\end{deluxetable}

In the limit of high loopgain, low parasitic resistances, and a negligible readout system transfer function, we have $\eta_{\mathrm{fmux}} \approx \eta_{\mathrm{true}}$.
However, the SPT-3G readout system has non-negligible parasitic series impedances at lower bias frequencies.
These parasitics generate a slight current bias, boosting the responsivity of the 95 and 150 \si{\giga\hertz} detectors.
The \SI{220}{\giga\hertz} detectors are operated at higher bias frequencies, where this series impedance is smaller \citep{Montgomery:2021}.
They therefore have a lower total responsivity, but also provide a means for roughly normalizing the excess responsivity of the 95 and 150 \si{\giga\hertz} detectors.

Although it is challenging to accurately estimate the magnitude of these systematic effects from our incomplete knowledge of the circuit dynamics, we have attempted to remove them using their correlation with the readout system bias frequencies.
In other words, we argue that $\eta_{\mathrm{true}}$ should be uncorrelated with bias frequency, apart from the two discontinuities associated with observing band changes (detector bias frequencies are grouped by observing band, which we do expect to have different values of $\eta_{\mathrm{true}}$).
For every wafer, we fit a function encapsulating the bias-frequency dependence of $\eta_{\mathrm{fmux}}$ and remove it, thereby removing all systematic readout calibration and responsivity effects dependent on bias frequency, up to a single overall scaling factor for all three bands.
In light of the fact that the \SI{220}{\giga\hertz} detectors are the least affected by this systematic, this overall scaling factor is constrained so that the median \SI{220}{\giga\hertz} $\eta$ value for each wafer is preserved pre- and post-correction.
On average, this correction process results in a \SI{29}{\percent} and \SI{15}{\percent} downward shift in the estimated cumulative efficiencies for the 95 and 150 \si{\giga\hertz} channels, respectively.

In \autoref{tab:optical_efficiencies}, we report the median $\eta$ measurement for each wafer and the full array.
The variation of median values between wafers is likely dominated by uncertainty associated with the median $\eta_{\mathrm{fmux}}$ \SI{220}{\giga\hertz} value, to which each wafer's corrected $\eta$ is scaled.
Considering the full array, the 95 and 150 \si{\giga\hertz} efficiencies agree with predicted efficiencies based on the optical model and known element properties of the instrument, whereas the \SI{220}{\giga\hertz} channels underperform predictions.

The consistency of measured and predicted optical loading in \autoref{sec:cryogenics} suggests that the level of unaccounted scattering or reflection through the system should not be unexpectedly high.
Our current conclusion is that our laboratory measurements of dielectric loss through alumina underestimated \SI{220}{\giga\hertz} signal attenuation through our multiple alumina elements, and that the details of the manufacturing process affected the optical properties of the final elements.
It is plausible that the optical properties of the larger monolithic elements may differ from those of the smaller samples explored during laboratory testing, despite nominally being the same formulation from the same vendor.

\begin{figure}[ht!]
\vspace{2mm}
\includegraphics[trim = 10 0 20 20, width=8.6cm]{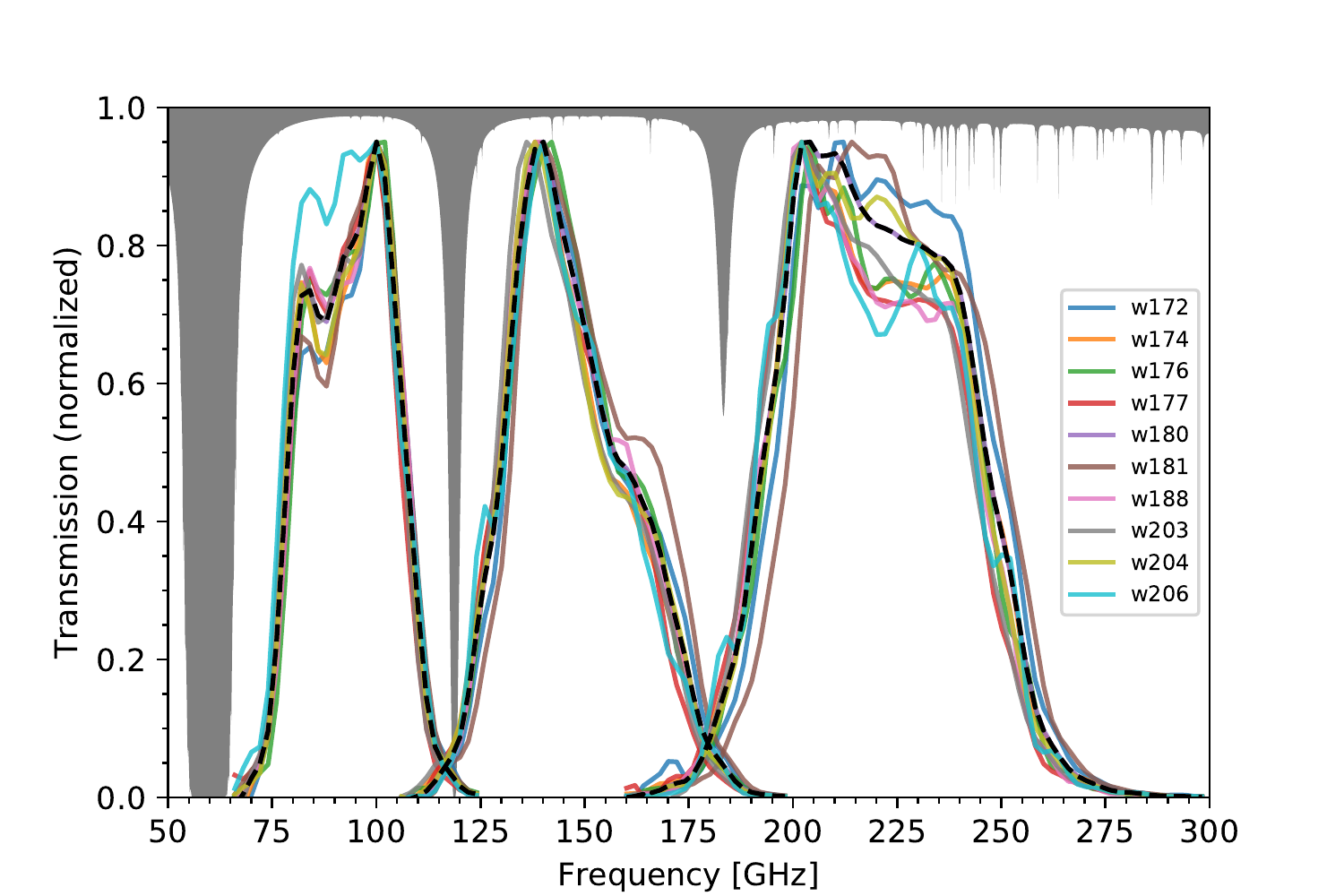}
\caption{
Frequency response, $g(\nu)$, of the SPT-3G receiver at \SIlist{95;150;220}{\giga\hertz} to a beam-filling, flat-spectrum source.
The solid lines show the response averaged by detector wafer, and the dashed line is the average response across all ten wafers.
The gray shaded region indicates atmospheric absorption for \SI{0.25}{\milli\meter} precipitable water vapor at the South Pole.
The frequency response has been arbitrarily normalized.  
The power on the detector, $P$, for a source with a spectrum, $I(\nu)$, would be $P = \int \eta_g \, g(\nu) I(\nu) d\nu$, where $\eta_g$ is the optical efficiency.
}
\label{fig:fts_bands}
\end{figure}

\subsection{Spectral Response} \label{subsec:spectral_response}

The SPT-3G receiver spectral response, averaged across the detector array and for each detector wafer, is shown in \autoref{fig:fts_bands}.
The array-averaged band center and width are summarized in \autoref{tab:integrated_instrument_parameters}.
The spectral response was measured in situ with a compact Fourier transform spectrometer (FTS), described in \citet{2018JLTP..193..305P, 2019ApOpt..58.6257P}.  
The measurement gives the SPT-3G receiver response to a beam-filling, flat-spectrum source at the receiver's vacuum window.
Therefore, the shape of the response is due to the combined transmission of the on-wafer triplexer filter, and the optical elements internal to the receiver, including any anti-reflection coatings.

\subsection{Beams and Absolute Temperature Calibration} \label{subsec:beams}

\begin{figure*}[ht!]
\includegraphics[scale=0.55]{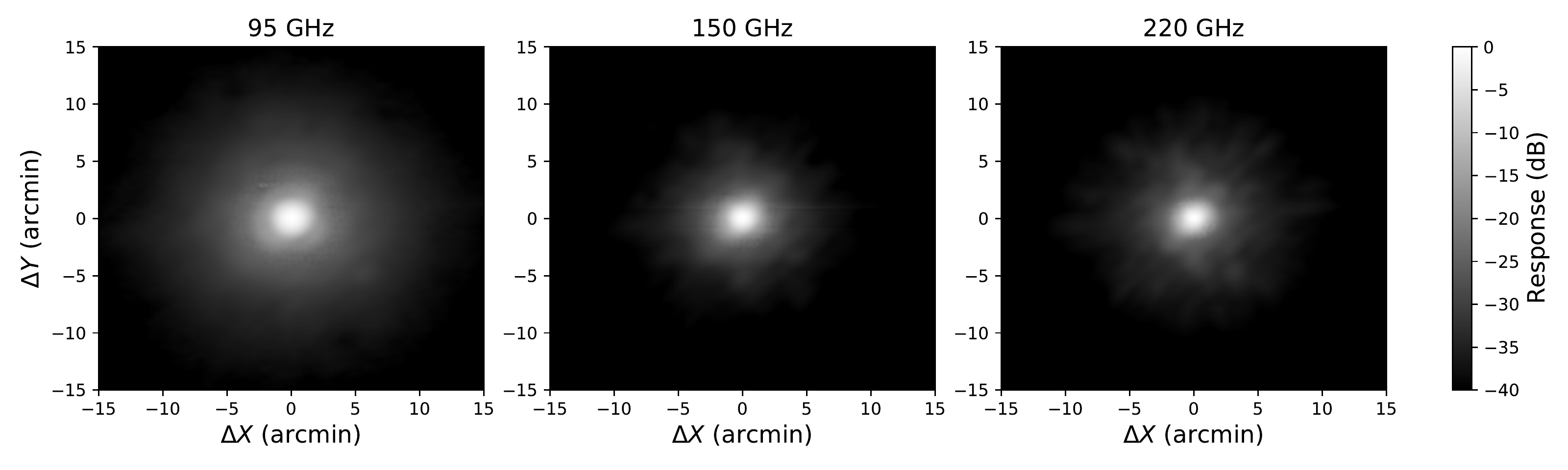}
\caption{
Composite beam response by band formed by stitching together point sources in the science observations with deep raster scans over the planets.
}
\label{fig:beams}
\end{figure*}

\begin{figure}[ht!]
\includegraphics[scale=0.55]{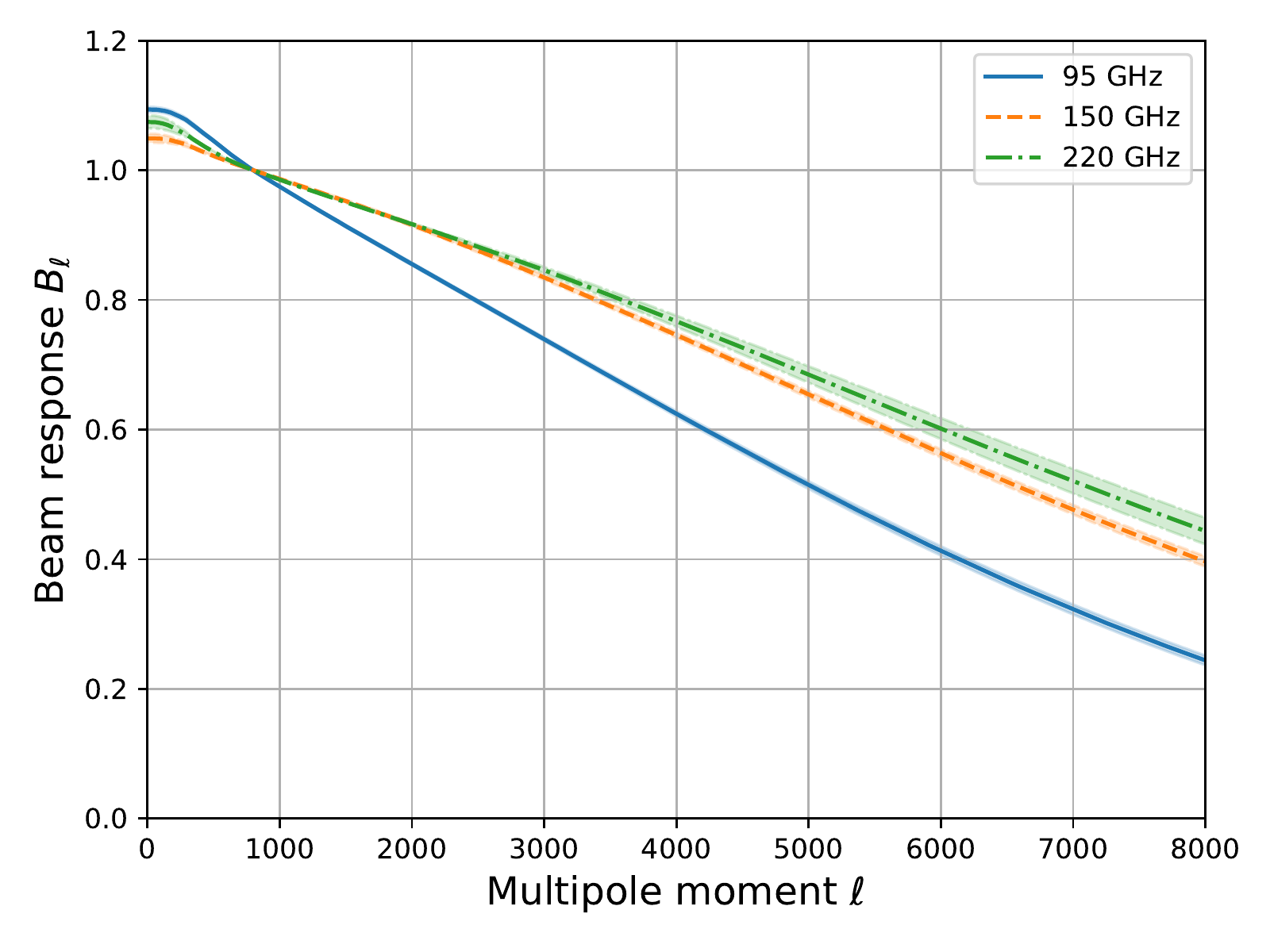}
\caption{
Composite beam response as a function of multipole moment $\ell$.
The shaded area shows the diagonal of the beam covariance calculated by repeating the computation with different subsets of planet and field source inputs and sampling from the covariance of the stitching parameters.
All beam response curves are normalized to unity at $\ell=800$.
}
\label{fig:b_ell}
\end{figure}

\begin{figure*}[ht!]
\begin{center}
\includegraphics[width=0.95\textwidth]{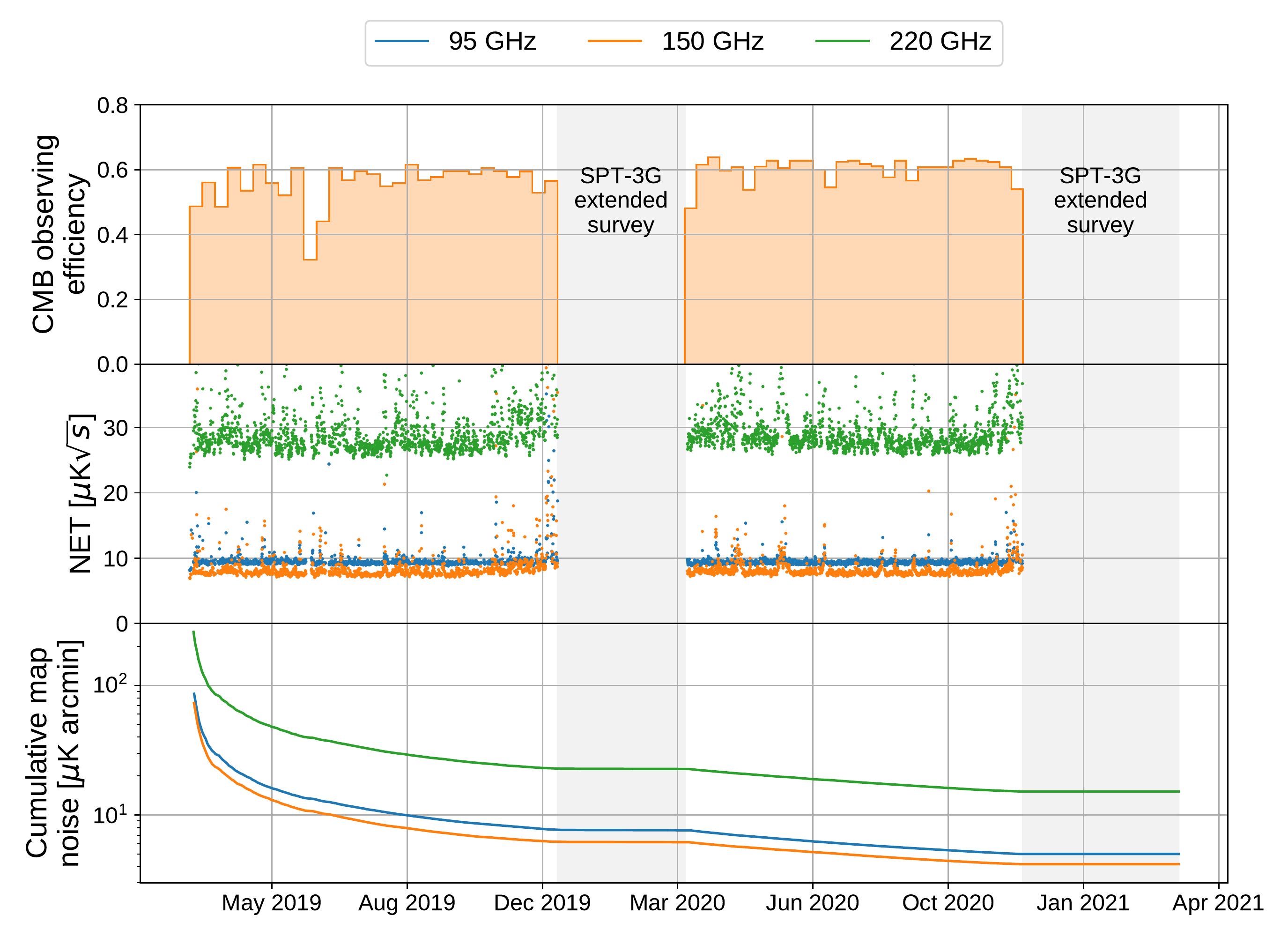}
\end{center}
\caption{
CMB observing efficiency, NET, and integrated noise of the 2019 and 2020 data from SPT-3G.
\emph{Top:} Fraction of total calendar time during which SPT-3G observed its main CMB fields. Time not spent observing the CMB was primarily spent tuning detectors, performing calibration observations, and cycling the sorption refrigerator.
\emph{Middle:} Array NET inferred from maps of subfield observations in the multipole range $\ell \in [3000,5000]$ for the \SI{95}{\giga\hertz} \emph{(blue)}, \SI{150}{\giga\hertz} \emph{(orange)}, and \SI{220}{\giga\hertz} \emph{(green)} detectors.
\emph{Bottom:} Depth in temperature of cumulative maps as a function of time.
Noise is estimated from angular multipoles of $\ell \in [3000,5000]$, averaged over the four subfields comprising the SPT-3G main survey.
The three curves correspond to the noise of the \SI{95}{\giga\hertz} \emph{(blue)}, \SI{150}{\giga\hertz} \emph{(orange)}, and \SI{220}{\giga\hertz} \emph{(green)} maps.
}
\label{fig:noise_livetime}
\end{figure*}

The SPT-3G point source response (the ``beam" response) is estimated in a hybrid manner similar to the SPT-SZ beam response~\citep{2011ApJ...743...28K} using a combination of point sources in the CMB observations and dedicated raster scans of Saturn.
These raster observations produce a high signal-to-noise (S/N) measurement of the beam response out to radii of tens of arcminutes, however, the detectors show evidence of non-linearity and saturation in the form of a suppressed response directly on the planet and a slow decay on the falling edge.
The response to point sources in the survey data is significantly more linear;
however, the available S/N is insufficient to resolve the extended beam structure.
The composite beam shape is calculated by stitching together the planet maps and point sources using a radial range where (a) the beam is resolved at high signal-to-noise in the point-source maps, (b) and the Saturn data can be cleaned of detector saturation.

The Saturn beam response is cleaned of saturation contamination by masking the data within a scan after the detector has come within one beam radius of the planet center until the end of the scan.
This method allows each detector to measure the rising edge of the beam and recover from the saturation before approaching Saturn from the other side.
This saturation masking radius is set by the extent of artifacts in the difference between the planet maps and the point source maps.
The CMB temperature anisotropies are resolved at high S/N in our planet raster scans.
We subtract the scanned Planck PR3 maps\footnote{\url{https://pla.esac.esa.int/}} from our planet maps, resulting in a percent-level change in our beam measurement at the largest angular scales.
These cleaned planet maps are fit to the field sources and blended to form a composite beam response, shown in \autoref{fig:beams}.
After dividing out the small contribution to the beam maps from statistical error in the pointing reconstruction and the finite Saturn disk size, the main lobe of the full detector array effective beam response can be approximated as a Gaussian with a FWHM of $1.57 / 1.17 / 1.04$ \si{\arcmin} at $95 / 150 / 220$ \si{\giga\hertz} (\autoref{tab:integrated_instrument_parameters}).

We perform the same beam calibration using four subsets of the field sources, split by the subfields defined in \autoref{subsec:cmb_fields}, using only one of the two deep planet raster scans, and sampling from the covariance of the amplitude and offset parameters used to align the field source and planet maps.
We find a statistical uncertainty on our $B_\ell$ measurement, shown in \autoref{fig:b_ell}, across the $\ell$ range of the SPT-3G science results.

We also measure the beam response in cross correlation with the Planck PR3 maps by comparing the temperature auto spectrum of our data with the cross spectrum of our data and the nearest frequency Planck map.
This analysis, described with further detail in \citet{2021arXiv210101684D}, provides a method for setting the absolute temperature calibration of SPT-3G coadded maps in each observing band.
In addition, this analysis allows us to estimate the beam window function $B_\ell$ in harmonic space, independent of the main position space analysis.
The harmonic space beam measurement is limited above $\ell \gtrsim 1500$ due to the Planck noise and beam size and below $\ell \lesssim 100$ by the timestream filtering used primarily to reduce atmospheric noise.
The uncertainties in this harmonic space beam calibration are established by direct Monte Carlo simulation.
We find consistent results with the measured position-space beam analysis over the range of multipoles where the temperature cross spectrum is informative.


\subsection{Sensitivity} \label{subsec:sensitivity}

\begin{figure*}[ht!]
\begin{center}
\includegraphics[width=0.95\textwidth]{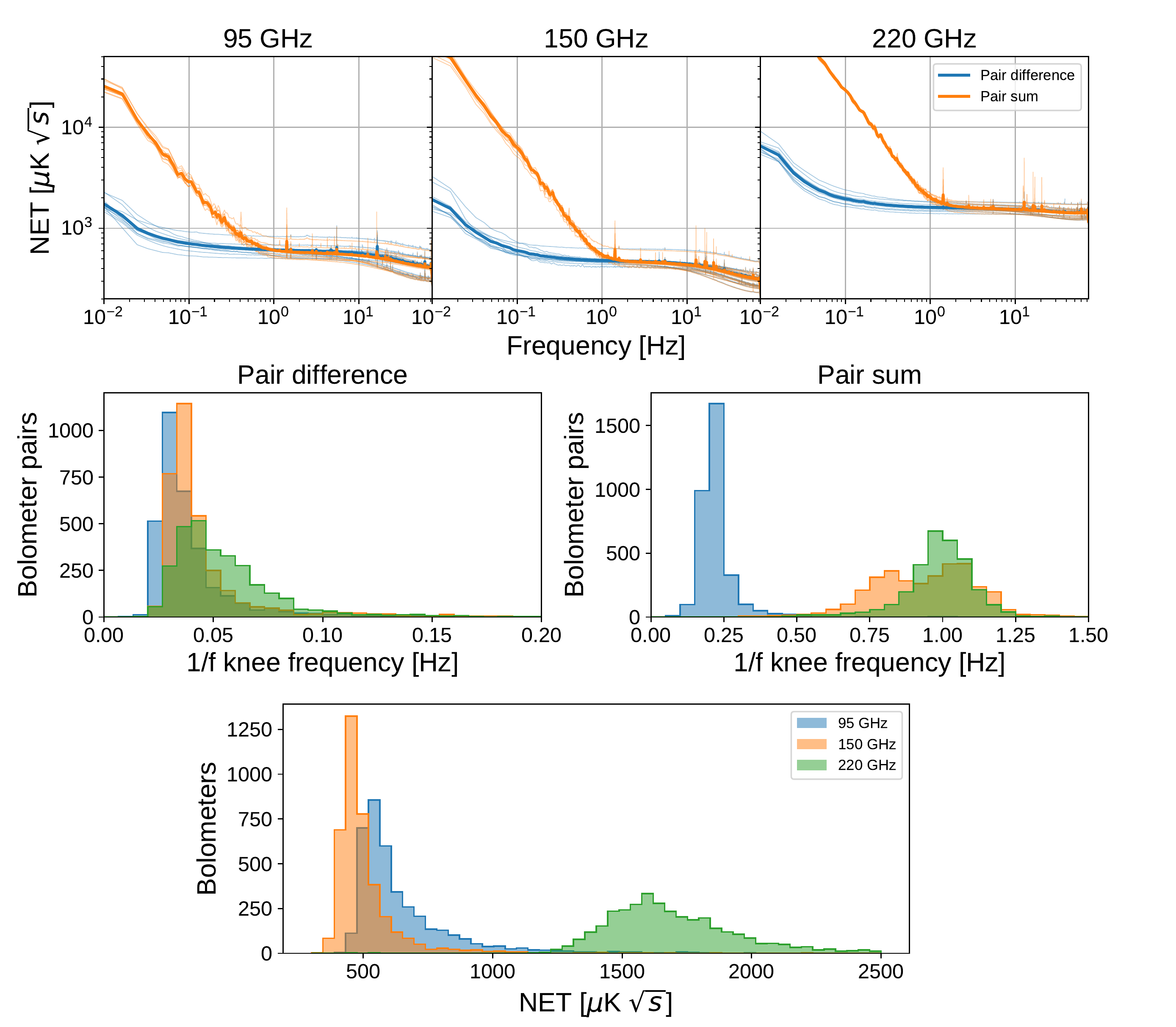}
\end{center}
\caption{
\emph{Top panel:} Mean noise amplitude spectral density for polarization pair sum \emph{(orange)} and difference \emph{(blue)} for 95 \emph{(left)}, 150 \emph{(middle)}, and \SI{220}{\giga\hertz} \emph{(right)} bolometers for the best \SI{90}{\percent} of noise measurements taken during the 2019 season.
Noise observations are taken with the telescope at rest and detectors operating as in CMB observations.
Thin lines are averages for individual wafers, while thick lines include all detectors.
\emph{Middle panel:} Median 1/f knees for the pair difference \emph{(left)} and pair sum \emph{(right)} noise spectra from noise observations.
Each entry in the histogram corresponds to the knee for a single polarization pair, with the median taken over all noise measurements in 2019.
\emph{Bottom panel:} Median NET for all bolometers. Each entry in the histogram corresponds to the noise level for a single bolometer, evaluated between 3 and 5 \si{\hertz}, with the median taken over all noise measurements in 2019. 
}
\label{fig:low_f_noise}
\end{figure*}

Over the entire 2019 and 2020 winter observing seasons (2019~March~21 to 2019~December~18, and 2020~March~21 to 2020~November~26), the fraction of time spent observing the CMB was \SI{58.4}{\percent}\footnote{An additional \SI{3.3}{\percent} of time spent observing the CMB is removed because the telescope is changing direction in azimuth between consecutive scans and is not moving at constant angular velocity.}, illustrated in \autoref{fig:noise_livetime}.
Unscheduled maintenance on the telescope drive system during June 2019 and several brief power outages account for most of the deviation from the optimal \SI{60}{\percent} efficiency.
The primary routine losses of observing time are due to cycling the $^3$He-$^3$He-$^4$He sorption refrigerator and re-tuning detectors (together, \SI{24}{\percent} loss) and calibration observations (\SI{12}{\percent} loss).

The median NET for the array during the 2019 and 2020 observing seasons was $9 / 8 / 28$ \si{\micro\kelvin~\sqrt{s}} at $95 / 150 / 220$ \si{\giga\hertz}, and the NETs per observation are shown in \autoref{fig:noise_livetime}.
To estimate these values, for each observation, we make a map with filtering settings optimized for detection of point sources and galaxy clusters, including a high-pass filter with a cutoff of $\ell = 500$. 
We then measure the autospectrum of the resulting map in the interval $\ell \in [3000,5000]$, well above the $1/\ell$ knee of the data.
The autospectrum includes both noise and sky signal, but at these multipoles, the sky signal in a single observation is negligible compared with the noise.
This map noise is converted to NET using the area of the uniform-coverage portion of the field and the time spent observing it.
These maps are created daily, primarily for data-quality monitoring purposes;
as such, they do not correct for the filter transfer function, biasing our reported NET to lower values by an $\ell$-dependent factor of 5--\SI{10}{\percent} in the range of $\ell \in [3000,5000]$.
In addition, the temperature calibration for these maps is derived from the HII-regions RCW38 and MAT5a, as described in \autoref{subsec:calibration}, which results in an additional $\lesssim$\SI{10}{\percent} difference relative to calibrating in cross correlation with Planck.
With these caveats, the coadd of these per-observation maps from the 2019 and 2020 winter seasons has a noise level of $5 / 4 / 15$ \si{\micro\kelvin~\arcmin} at $95 / 150 / 220$ \si{\giga\hertz} in temperature (\autoref{tab:integrated_instrument_parameters}).
We plan to continue observing the same sky area with SPT-3G for the next three austral winters (through the end of 2023).
Since no major changes to the instrumental configuration or observing strategy are planned, each of these upcoming seasons should have sensitivity comparable to the level achieved during 2019 and 2020.

At frequencies below \SI{1}{\hertz}, atmospheric temperature fluctuations result in a significant increase in noise above the white photon-noise floor. 
Since the atmospheric signal is largely unpolarized, these fluctuations can be efficiently removed by differencing timestreams from detectors with orthogonal polarizations in the same pixel.
\autoref{fig:low_f_noise} shows the reduction in low-frequency atmospheric noise achieved by differencing polarization pairs, with spectra measured with the telescope stationary and detectors operating as in CMB observations.
Quantitative characterization of low-frequency noise in the map domain is an area of ongoing study and will be described in future publications that use these data products, while the low-frequency noise performance of the readout electronics is discussed in \citet{2020JLTP..199..182B}.

\added{In situ measurements of detector-level NETs across the array are shown in the bottom panel of \autoref{fig:low_f_noise}.
Using our understanding of the SPT-3G optical model, detector properties, and readout system design, we are able to roughly predict (with some caveats) expected detector-level NETs.
In comparing these values, we find acceptable agreement between our measurements and noise model at 95 and 150 \si{\giga\hertz}, but elevated noise levels among the \SI{220}{\giga\hertz} detectors.
The higher-than-expected \SI{220}{\giga\hertz} NET levels are very likely caused by the instrument's lower-than-expected cumulative optical efficiency across the \SI{220}{\giga\hertz} band~(\autoref{subsec:optical_efficiency}), as well as by elevated readout system noise at the higher bias frequencies~(\autoref{subsec:noise}).}


\begin{deluxetable*}{lccc}
\tablecaption{
Measured Array-Averaged Instrument Parameters.
}
\tablewidth{0pt}
\tablehead{
\colhead{} & \colhead{\SI{95}{\giga\hertz}} & \colhead{\SI{150}{\giga\hertz}} &  \colhead{\SI{220}{\giga\hertz}} 
}
\startdata
Measured Optical Efficiency & $ 0.25 \pm 0.07 $ & $ 0.44 \pm 0.14 $ & $ 0.13 \pm 0.04 $ \\
$\phn\phn$(excluding stop efficiency) & 0.45 & 0.54 & 0.14 \\
\hline
Band Center (\si{\giga\hertz}) & $93.8 \pm 0.7$ & $147.0 \pm 1.2$ & $219.9 \pm 2.0$\\
Band Width (\si{\giga\hertz}) & $26.4 \pm 1.2$ & $32.5 \pm 0.7$ & $53.6 \pm 1.9$\\
\hline
Beam FWHM (\si{\arcmin}) & 1.57 & 1.17 & 1.04 \\
\hline
Median NET (\si{\micro\kelvin \sqrt{s}}) & 9 & 8 & 28 \\
$T$ Map Depth (2019+2020), & 5 & 4 & 15 \\
$3000<\ell<5000$ (\si{\micro\kelvin\arcmin}) \\
\hline
Polarization-Angle Uncertainty (\si{\deg}) & 2.0 & 2.2 & 4.5
\enddata
\tablecomments{
In addition to the measured instrument optical efficiency taken from \autoref{subsec:optical_efficiency}, we quote the same efficiency measurement after excluding the predicted Lyot stop spillover loss, as described in \autoref{tab:bolometer_loading}.
The band center is defined as $\int \nu g(\nu) d\nu / \int g(\nu) d\nu$ and band width defined as $\int g(\nu) d\nu$, where $g(\nu)$ is the SPT-3G receiver frequency response (normalized to 1) to a beam-filling, flat-spectrum source (as shown in \autoref{subsec:spectral_response}).
The uncertainties reflect the spread across detector wafers.
}
\label{tab:integrated_instrument_parameters}
\end{deluxetable*}


\subsection{Polarization Calibration} \label{subsec:polcal}

As described in \autoref{subsec:pixel_design}, each detector wafer contains dual-polarization pixels with alternating \SI{45}{\deg} rotation to measure the $Q$ and $U$ Stokes parameters.
Given the hexagonal shape of the wafers, each containing detectors with four polarization orientations, there are a total of twelve polarization orientations over the entire detector array, each separated by \SI{15}{deg}.
Each detector is attributed a nominal polarization angle based on the wafer's orientation during installation.
The goal of our polarization calibration procedure is to confirm that each detector has been assigned the correct nominal polarization angle,\footnote{This mapping is not known unambiguously a priori for every detector because of frequency scatter and imperfect yield in the LCR resonators that comprise the multiplexing readout circuit.} which we then assume to be its true polarization angle when constructing CMB maps.
This assumption results in a modest loss of polarization efficiency, which can likely be improved by further assigning average polarization angles to subsets of detectors that have a systematic polarization rotation relative to the nominal angle (e.g. pixels with left- and right-handed sinuous antennas).

Because of the difficulty in performing polarization calibration with a terrestrial source in the far-field of a large-aperture CMB telescope, we use dedicated observations of Centaurus A (CenA, NGC 5128) to derive the relative polarization angles of detectors by fitting its large, arcminute-scale, polarized radio lobes.
The CenA method is described here and is currently being studied using SPT-3G data. 
Final results and further discussion of this method, as well as of a similar method using the polarization of the CMB itself, will be included in a future paper.

Each CenA observation takes $\sim$3 hours and consists of a dense, \SI{0.25}{\arcmin} raster over a \SI{3}{\degree} $\times$ \SI{3}{\degree} area, centered on CenA.
After the observation, a full-array, polarized coadded map is constructed for each observing band, along with single-detector maps for each detector in the array.
Each single-detector map, with observing band $b$, is then fit to a model template given by
\begin{equation}\label{eq:polarizationfit}
t_b = g {\big [} (2-\rho)T_b + \rho \cos(2\Theta)Q_b + \rho \sin(2\Theta)U_b {\big ]},
\end{equation}
where $T_b$, $Q_b$, and $U_b$ are the observed, per-band detector-array coadded maps; an overall gain ($g$), polarization angle ($\Theta$), and polarization efficiency ($\rho$) per detector are the free parameters of the fit.
Because the template coadd maps are constructed using the nominal polarization angles, efficiencies, and detector gains, we lose the ability to constrain the absolute global parameters.
The fit parameters for each detector are estimated from a least-squares fit of each single-bolometer map to the model template of \autoref{eq:polarizationfit}.
Since a single observation of CenA by a single bolometer does not provide enough sensitivity to yield a meaningful measurement of its polarization angle, this calibration was performed on a cadence of approximately once per week during the 2019 austral winter.

The polarization angle for each detector was computed for each of 30 observations of CenA during the 2019 season.
The mean polarization angle and associated uncertainty for each detector was estimated by the sample mean and standard error over the 30 calibration observations.
With this method, we measure every detector's polarization angle with a median uncertainty of $2.0 / 2.2 / 4.5$~\si{\deg} for the $95 / 150 / 220$~\si{\giga\hertz} detectors (\autoref{tab:integrated_instrument_parameters}).
This uncertainty is well below the \SI{15}{\deg} difference between nominal polarization angles, allowing us to assign each detector to its nominal angle with high confidence.
These nominal polarization angles were assumed in the construction of maps for recent analyses of SPT-3G data (e.g. \citealt{2021arXiv210101684D}), as detailed characterization of any real deviations from these nominal angles is a topic of ongoing study.
While these measurements of individual detector polarization angles do not imply significant deviation from nominal angles, it may be possible to correct for deviations from nominal angles by averaging many detectors across the array or a wafer.

At the level of sensitivity expected for SPT-3G, the primary impact of assuming nominal polarization angles during mapmaking is a small decrease in the average polarization efficiency of the experiment.
One notable cause of this decrease is the polarization wobble of the sinuous antenna---which we detect at high significance in our analysis of CenA observations---together with our choice to use an equal number of pixels with left-handed and right-handed antennas (\autoref{subsec:pixel_design}).
Left-handed and right-handed pixels have polarization angles that are slightly offset by equal and opposite magnitudes from the nominal angle, resulting in decreased overall polarization efficiency.
In \cite{2021arXiv210101684D}, we corrected for the change in polarization calibration by comparing the $TE$ and $EE$ power spectra of SPT-3G to those of Planck, implying polarization efficiencies of $97.2 / 94.6 / 88.0$~percent for $95 / 150 / 220$~\si{\giga\hertz} maps.
This comparison indicates that even without correcting for the mean polarization rotation due to the polarization wobble, the assumption of nominal angles has at most a modest effect on the overall polarization sensitivity.
We anticipate an even smaller efficiency loss in maps constructed using the mean measured polarization angles for left-handed and right-handed detectors at each orientation.

\section{CONCLUSION} \label{sec:conclusion}

We have presented the design and integrated performance of the SPT-3G instrument, which has already achieved temperature map-depth of $5 / 4 / 15$ \si{\micro\kelvin~\arcmin} at $95 / 150 / 220$ \si{\giga\hertz}, using two years of data from a multiyear survey.
SPT-3G is currently observing and plans to continue doing so through the end of the 2023 season, providing deep, arcminute-scale resolution CMB maps that will be useful for a wide range of scientific analyses.

Already, measurements of the $EE$ and $TE$ power spectra using four months of SPT-3G data in 2018 (half of a typical observing season) have improved upon previous results from SPTpol at multipoles $\ell \lesssim 1500$~\citep{2020JLTP..199..182B}, and provided stronger constraints on extensions to the $\Lambda$CDM cosmological model~\citep{2021arXiv210313618B}.
The instrument's increased detector count has improved instantaneous sensitivity to small angular-scale features, allowing for near real-time detection of Galactic and extragalactic millimeter-wave transient sources~\citep{2021arXiv210306166G}.
Similarly, SPT-3G's large observing footprint and high re-observation cadence provides a powerful look into high-resolution time-domain astrophysics of a large range of sources from blazars to low-luminosity AGN and flaring stars.

In addition, analyses are underway to use current data to measure temperature, polarization, and lensing power spectra on arcminute scales to further explore tensions with the $\Lambda$CDM model and constrain possible extensions.
Joint efforts using SPT-3G data along with BICEP/\emph{Keck} data to de-lens the $B$-mode polarization power spectrum will provide unprecedented constraints on the energy scale of inflation.
SPT-3G maps are also currently being used to expand catalogs of emissive point sources and high-redshift galaxy clusters. 
The complete SPT-3G survey will produce maps with an unprecedented combination of sensitivity and resolution that will enable significant advances in millimeter-wave astronomy and cosmological constraints from the CMB.


\section*{acknowledgments}

The South Pole Telescope program is supported by the National Science Foundation (NSF) through grants PLR-1248097 and OPP-1852617.
Partial support is also provided by the NSF Physics Frontier Center grant PHY-1125897 to the Kavli Institute of Cosmological Physics at the University of Chicago and the Kavli Foundation.
Argonne National Laboratory's work was supported by the U.S. Department of Energy, Office of High Energy Physics, under contract DE-AC02-06CH11357.  This work was performed, in part, at the Center for Nanoscale Materials, a U.S. Department of Energy Office of Science User Facility, and supported by the U.S. Department of Energy, Office of Science, under Contract No. DE-AC02-06CH11357. 
We acknowledge R. Divan, L. Stan, C.S. Miller, and V. Kutepova for supporting our work in the Argonne Center for Nanoscale Materials.
Work at Fermi National Accelerator Laboratory, a DOE-OS, HEP User Facility managed by the Fermi Research Alliance, LLC, was supported under Contract No. DE-AC02-07CH11359.
NWH acknowledges support from NSF CAREER grant AST-0956135.
The McGill authors acknowledge funding from the Natural Sciences and Engineering Research Council of Canada, Canadian Institute for Advanced Research, and the Fonds de recherche du Qu\'ebec Nature et technologies.
This material is based upon work supported by the U.S. Department of Energy, Office of Science, Office of High Energy Physics under Award Number DE-SC-0015640.
MA and JV acknowledge support from the Center for AstroPhysical Surveys at the National Center for Supercomputing Applications in Urbana, IL.
JV acknowledges support from the Sloan Foundation.

%

\vspace{5mm}
\facilities{Amundsen--Scott South Pole Station}


\software{
          IPython \citep{ipython},
          LMFIT \citep{newville_matthew_2014_11813},
          Matplotlib \citep{matplotlib_package},
          NumPy \citep{numpy_package},
          Pandas \citep{mckinney-proc-scipy-2010},
          SciPy \citep{scipy_package}
          }


\bibliography{references}{}
\bibliographystyle{aasjournal}



\end{document}